\documentclass[amsmath,amssymb,12pt]{article}

\usepackage{graphicx}
\usepackage{amssymb}
\usepackage{epstopdf}
\usepackage{pdfsync}

\usepackage{color}

\usepackage{graphicx}
\usepackage{amsmath}

\input{tcilatex}

\usepackage{amsfonts}
\usepackage{mathrsfs}

\usepackage{makeidx}

\makeindex

\begin{document}
 
  \title{Quantum Control Theory\thanks{This research was supported by the Australian Research Council Centre of Excellence for Quantum Computation and Communication Technology (project number CE110001027), and US Air Force Office of Scientific Research Grant FA2386-12-1-4075.}
\author{M.R.~James\thanks{ARC Centre for Quantum Computation and Communication Technology, Research School of Engineering, Australian
National University, Canberra, ACT 0200, Australia (e-mail: 
Matthew.James@anu.edu.au)}
  }}
 \date{\today}

\maketitle

\begin{abstract}
This paper explains some fundamental ideas of {\em feedback} control of quantum systems through the study of a relatively simple two-level system coupled to optical field channels. The model for this system  includes both continuous and impulsive dynamics. Topics covered in this paper include open and closed loop control, impulsive control, optimal control, quantum filtering, quantum feedback networks, and coherent feedback control. 

\end{abstract}

\tableofcontents

\section{Introduction}
\label{sec:intro}

We are currently witnessing the beginnings  of a new era of technology---an era of technology that fully exploits the unique resources of quantum mechanics. This \lq\lq{second quantum revolution}\rq\rq \ (Dowling and Milburn, 2003) will demand a new generation of  engineering methodologies. {\em Quantum engineering} will be a new branch of engineering focused on
the design and manufacturing of quantum technologies. At present, quantum engineering is embryonic and includes activities that
range from laboratory experimentation 
of new devices and systems, to the development of the theory that inspires and supports the creation of these new devices and systems. 
Just as today's non-quantum  engineering arose from  the foundations of classical physics, mathematics, and (most importantly) entrepreneurship, quantum engineering is beginning to evolve from  the fundamental principles of  quantum physics and mathematics and examples of quantum entrepreneurship can now be seen.

This document is focused on one aspect of this second quantum revolution---quantum control.  Control, of course, is essential  to technology, and indeed played a key enabling role during the industrial revolution. Quantum control originated in the sciences, and we are now beginning to see the growth of quantum control research in engineering.  In what follows we look at some recent developments in quantum control theory from the perspective of a control theorist. We
 hope that readers will see how various contributions to quantum control fit together. In particular, the paper explains some fundamental ideas of {\em feedback} control through the study of a relatively simple two-level system coupled to optical field channels. This model for this system  includes both continuous and impulsive dynamics. 

This paper is organized as follows. 
In  Section \ref{sec:types} we describe some of the main types of quantum control that have appeared in the literature. In Section \ref{sec:open-loop} we discuss several aspects of open loop control: (i) time optimal control for a closed system, (ii) impulsive optimal control, and (iii) regulation of systems subject to relaxation. In preparation for our more detailed discussion of quantum feedback, Section \ref{sec:networks} describes  {\em quantum feedback networks} (QFN), while Section \ref{sec:filter} presents some of the basic ideas of {\em quantum filtering}, which  is seen as a natural  extension of statistical reasoning to  quantum mechanics.   We then consider two types of optimal measurement feedback
   control problems in Section \ref{sec:optimal}, and discuss the important idea of {\em information states} for feedback control. 
   Finally, Section \ref{sec:coherent} presents some ideas concerning coherent feedback control. 
 
 \subsection*{Notation and Preliminaries}

 Quantum mechanics is usually represented mathematically using a Hilbert space. In this chapter,  $\mathfrak{H}$ will denote a finite dimensional Hilbert space, say $\mathfrak{H}=\mathbb{C}^n$, the $n$-dimensional complex vector space.
 In Dirac's notation, the  inner product for $\mathfrak{H}$ is denoted
$$
\langle \psi \vert \phi \rangle =  \sum_{j=1}^n \psi^\ast_j \phi_j .
$$
 The    vector $\vert \phi \rangle \in \mathfrak{H}$  is denoted (represented by a column vector of length $n$ with complex entries $\psi_j$), is called a {\em ket}, while dual (row) vectors   are called {\em bras} and written  as
$$
  \langle \psi \vert .
$$

A linear operator on $\mathfrak{H}$ is denoted $A$\footnote{We do not use $\hat{}$ to indicate operators. Later, we will use $\hat{}$ to denote an {\em estimate}  $\hat X$ of operators  $X$.} (typically represented by an $n\times n$ complex matrix). 
For any operator $A$ its {\em adjoint}\footnote{Note that we use $A^\ast$ to denote {\em adjoint} of an operator $A$ instead of $A^\dagger$.  However, if $A=(a_{jk})$ is a matrix (with operator or complex number entries $a_{jk}$), we write $A^\dagger = (a^\ast_{kj})$ (conjugate transpose).}   
$A^\ast$ is an operator defined by
$$
\langle A^\ast \psi \vert \phi \rangle = \langle \psi \vert A\phi \rangle \  \ \text{for all} \ \langle  \psi \vert,
\vert \phi \rangle .
$$
The  adjoint  $\vert \psi \rangle^\ast$ of a vector $\vert \psi \rangle$ is a dual vector (represented by a row vector):
$$
\langle \psi \vert = \vert \psi \rangle^\ast .
$$
\index{normal}  \index{self-adjoint}  \index{unitary}
An operator $A$  is called {\em normal} if
$AA^\ast=A^\ast A$. Two important types of normal operators
are  {\em self-adjoint} ($A=A^\ast$), and {\em unitary}
($A^\ast=A^{-1}$).

\index{spectral theorem}
The {\em spectral theorem} says that 
 if $A$ is a self-adjoint operator on a finite dimensional Hilbert space $\mathfrak{H}=\mathbb{C}^n$, 
the eigenvalues (not necessarily distinct) $\mathrm{spec}(A)=\{ a_j \}_{j=1}^n$ of $A$ are real  and  $A$ can be written as
\begin{equation}
A = \sum_{a\in \mathrm{spec}(A) } a  P_a ,
\label{spectral-1}
\end{equation}
where $P_a$ is the projection 
\begin{equation}
P_a = \sum_{j : a_j= a} \vert a_j \rangle  \langle a_j \vert ,
\label{spectral-1a}
\end{equation}
and $\{ \vert a_j \rangle \}_{j=1}^n$ are the corresponding orthonormal eigenvectors.
The projections resolve the identity $\sum_{a \in  \mathrm{spec}(A)} P_a = I$.

Let's denote the collection of all (bounded, linear) operators on a Hilbert space $\mathfrak{H}$ by $\mathscr{B}=\mathscr{B}(\mathfrak{H})$.  The set
$\mathscr{B}$ can be thought of as a {\em vector space}, where operators $A \in \mathscr{B}$ are \lq\lq{vectors}\rq\rq. Indeed, if $\alpha_1$ and $\alpha_2$ are complex numbers and $A_1, A_2 \in \mathscr{B}$, then the linear combination $\alpha_1 A_1+\alpha_2 A_2$ is the operator in $\mathscr{B}$ defined by
\begin{equation}
(\alpha_1 A_1+\alpha_2 A_2) \vert \psi \rangle = \alpha_1 A_1 \vert \psi \rangle +\alpha_2 A_2 \vert \psi \rangle, \ \text{for all} \ \vert \psi \rangle \in \mathfrak{H}.
\label{eq:operators-linear-comb}
\end{equation}
Note that the linear combination on the RHS of (\ref{eq:operators-linear-comb}) is a linear combination of vectors in the Hilbert space $\mathfrak{H}$.
We can multiply operators,
\begin{equation}
 (A_1A_2) \vert \psi \rangle = A_1( A_2 \vert \psi \rangle ) , \ \text{for all} \ \vert \psi \rangle \in \mathfrak{H},
\label{eq:operators-mult-comb}
\end{equation}
so that $A_1A_2 \in \mathscr{B}$ if $A_1, A_2 \in \mathscr{B}$. Also, the adjoint $A^\ast \in \mathscr{B}$ if $A \in \mathscr{B}$. So the collection $\mathscr{B}$ of operators is closed under addition, scalar multiplication, multiplication, and adjoints---mathematically, $\mathscr{B}$ is called a {\em $\ast$-algebra}. This mathematical structure is fundamental to quantum mechanics.
\index{$\ast$-algebra}   \index{algebra!$\ast$}
 
 Tensor products are used to describe \emph{composite systems}. If
 $\mathcal{H}_1$ and $\mathcal{H}_2$ are Hilbert spaces, the {\em
 tensor product} $\mathcal{H}_1 \otimes \mathcal{H}_2$ is the Hilbert
 space consisting of linear combinations of the form $\vert \psi_1 \rangle \otimes
\vert  \psi_2 \rangle $, and inner product $\langle \psi_1 \otimes \psi_2, \phi_1
 \otimes \phi_2 \rangle = \langle \psi_1, \phi_1 \rangle \langle
 \psi_2, \phi_2 \rangle$. Here, $\vert \psi_1\rangle, \vert \phi_1 \rangle \in \mathcal{H}_1$ and
 $\vert \psi_2 \rangle, \vert \phi_2 \rangle \in \mathcal{H}_2$.  If $A_1$ and $A_2$ are operators
 on $\mathcal{H}_1$ and $\mathcal{H}_2$, respectively, then
 $A_1\otimes A_2$ is an operator on $\mathcal{H}_1 \otimes
 \mathcal{H}_2$ and is defined by $(A_1\otimes A_2)(\vert \psi_1 \rangle \otimes
\vert  \psi_2 \rangle) = A_1 \vert \psi_1 \rangle \otimes A_2 \vert \psi_2 \rangle$. Often, 
$\vert \psi_1 \rangle \otimes
\vert  \psi_2 \rangle $ is written as $\vert \psi_1 \psi_2 \rangle$, and 
$A_1\otimes A_2$ is
 written $A_1A_2$.

\section{Types of Quantum Control}
\label{sec:types}

\begin{figure}[h]
\begin{center}
\includegraphics[scale=0.8]{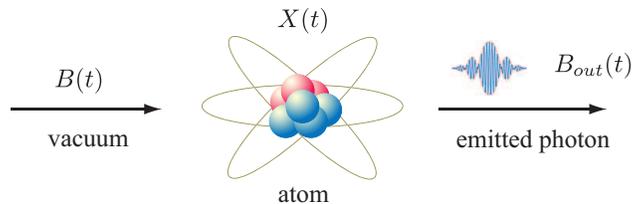}
\caption{An atom interacting with an electromagnetic field (e.g. photon emission).}
\label{fig:atom-field-1}
\end{center}
\end{figure}

Due to their relative simplicity and tractability,
the two-level quantum system and  the quantum harmonic oscillator are two of the most important prototype models for  quantum systems. These models  are widely used for describing  real physical systems,  as well as  for tutorial  purposes. In this article we discuss a range of aspects of quantum control primarily  focused on the basic two-level quantum system as the system to be controlled. Two level systems are used in quantum computing as the {\em qubit}, in NMR spectroscopy as a basic {\em spin system}, and in quantum optics as a model for an  {\em atom} with two energy levels. The oscillator plays a role in the representation of an electromagnetic field to which the two-level atomic system is coupled, Figure \ref{fig:atom-field-1}, \cite[Fig. 9.1]{GZ00}.

\begin{figure}[h]
\begin{center}
\includegraphics{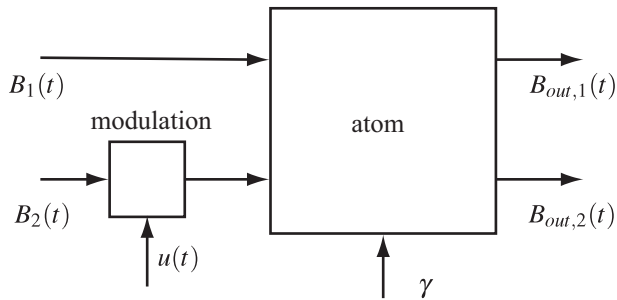}
\caption{Classical open-loop control of a  two-level atom showing input $B_1(t)$, $B_2(t)$  and output $B_{1,out}(t)$, $B_{2,out}(t)$  fields, as well as classical control variables $u(t)$, $\gamma$.}
\label{fig:atom-1}
\end{center}
\end{figure}

In this chapter we consider a two-level atom coupled to  several  electromagnetic  field channels.
In the absence of any other influences, the behavior of the atom  will evolve in time  according to the laws of quantum mechanics, as determined by the self-energy of the atom, the nature of the fields and how they are coupled to the atom. The field channels may be used to gather information about the atom, and to influence the behavior of the atom.
 Figure \ref{fig:atom-1} shows a schematic representation of an atom coupled to a pair of optical field channels, one of which is used to describe light shone on the atom, while the other channel contains the outgoing light, \cite[Sec. 9.2]{GZ00}. Also shown is a channel used to apply a rapid radio-frequency (RF) pulse. 
As will be explained, light shone on the atom  may be regarded as   a control signal that may be  classical or quantum  in nature; the later allowing for coherent feedback control.
 In this article we are interested in idealized impulsive models for pulses (zero width and infinite height).\footnote{In reality all pulses have non-zero width and finite height, and the impulsive model is useful when the time scales are such that the response to a rapid pulse is much faster than the other dynamics of the system.} The impulsive signals are classical signals, but may be used to effect coherent transformations of the system.


We now present   a model for a controlled  two level  atom.
Since a two-level atom is a quantum system with two energy levels,  the model  makes use  of the Pauli matrices
\begin{eqnarray}
\sigma_0 =  I = \left(  \begin{array}{cc}
1 & 0 \\
0 & 1
\end{array} \right) , \ \sigma_x =   \left(  \begin{array}{cc}
0 & 1 \\
1 & 0
\end{array} \right) , \
 \sigma_y =    \left(  \begin{array}{cc}
0 & -i \\
i & 0
\end{array} \right) , \ 
\sigma_z =   \left(  \begin{array}{cc}
1 & 0 \\
0 & -1
\end{array} \right) .
\label{eq:pauli-def}
\end{eqnarray}
Any observable of the atom can be expressed in terms of these matrices. 
 In particular, the atomic  energy levels are the eigenvalues $\pm \frac{1}{2} \omega$ of the Hamiltonian $H=\frac{1}{2}\omega \sigma_z$ describing the self-energy of the atom. 

The atom interacts with the field channels by an exchange of energy that may be described by first principles in terms of an interaction Hamiltonian, \cite[Chapter 3]{GZ00}. In this article we use an idealized  quantum noise model for the open atom-field system which is well justified theoretically and experimentally, \cite{HP84}, \cite{GC85}, \cite{KRP92}, \cite{GZ00}. Each  field channel has input and output components, modeled as quantum stochastic processes.  The input processes,   $B_1(t)$ and $B_2(t)$,  drive an interaction-picture equation for a unitary operator $U(t)$ governing the atom-field system.  If $\vert \psi_a \rangle$ and $\vert \psi_f \rangle$ are initial atomic and field states respectively, the state of the atom-field system at time $t$ is $U(t) \vert \psi_a \psi_f \rangle$.

This continuous (in time) stochastic unitary model holds in the absence of the above-mentioned impulsive actions. Now suppose that impulses are  applied at times $0 \leq \tau_0 < \tau_1 < \ldots$ by selection of a unitary $V$  from a set $\mathbf{V}$ of unitaries.  If $V_k \in \mathbf{V}$ is selected at time $\tau_k$, the state immediately after the impulse has been applied is $V_k U(\tau_k) \vert \psi_a \psi_f \rangle$. So the impulse is modeled as instantaneously effecting a unitary transformation. Combining the continuous and impulsive motions, we see that if $\tau_k < t < \tau_{k+1}$ the overall unitary at time $t$ is
\begin{equation}
U(t) = U(t, \tau_k) V_{k} U( \tau_k, \tau_{k-1}) V_{{k-1}} \ldots U(\tau_1, \tau_0)V_{0} U(\tau_0, 0) ,
\label{eq:hybrid-1}
\end{equation}
where $V_k \in \mathbf{V}$  indicates which impulse was selected at time $\tau_k$ and $U(t,s)$ is the unitary for the continuous motion on the time interval $(s,t)$ ($s < t$), with $U(t,t)=I$.

 Let's now look at the equations governing the hybrid continuous-impulsive dynamics. We suppose that the two-channel field is initially in the vacuum state, which we denote by $\vert \psi_f \rangle= \vert 00 \rangle$. In this case the input processes $B_1(t)$ and $B_2(t)$ are independent  quantum Wiener processes, for which the non-zero Ito product are $dB_j(t) dB^\ast_j(t)=dt$ ($j=1,2$). The atom-field coupling is determined by   coupling operators $L_1=\sqrt{\kappa_1}\, \sigma_-$ and $L_2=\sqrt{\kappa_2}\, \sigma_-$, where
 $$
 \sigma_- = \left(
 \begin{array}{cc}
 0 & 0
 \\
 1 & 0
 \end{array}
 \right)
 $$
is the lowering operator (the raising operator is defined by $\sigma_+ = \sigma_-^\dagger$), and $\kappa_1, \kappa_2 > 0$ are  parameters  describing the strength of the coupling to each channel.  
The evolution of the unitary $U(t)$ is given by
\begin{eqnarray}
d U(t) &=& \{ \sqrt{\kappa_1}\, dB_1^\ast(t) \sigma_-  - \sqrt{\kappa_1}\, \sigma_+ dB_1(t) 
+ \sqrt{\kappa_2}\, dB_2^\ast(t) \sigma_-  - \sqrt{\kappa_2}\, \sigma_+ dB_2(t) 
\nonumber \\ 
&& \hspace{2.0cm} 
- \frac{1}{2} ((\kappa_1+\kappa_2) \sigma_+ \sigma_-   + i  \omega \sigma_z   ) dt \} U(t),
\ \ \tau_k < t \leq \tau_{k+1},
\nonumber \\
U(\tau_k^+) & = & V_{k} U(\tau_k) .
\label{eq:hybrid-2}
\end{eqnarray}
Here, $\tau_k^+$ indicates the value immediately after  $\tau_k$, i.e., the limit from the right. Equation (\ref{eq:hybrid-2}) is {\em quantum stochastic differential equation  with impulses}, whose solution $U(t)$ (of the form  (\ref{eq:hybrid-1})) is determined by a sequence of time-impulse pairs
\begin{equation}
\gamma = ( (\tau_0, V_0), (\tau_1, V_1), \ldots ) .
\label{eq:hybrid-3}
\end{equation}

So  far, we have not described how modulation of the field can be used to control the atom. Let us now do so. Suppose we modulate the second field channel to be in a coherent state $\vert \frac{gu}{\sqrt{\kappa_2}}  \rangle$, where $u(\cdot)$ is a classical function of time (a classical control signal).  We may model this by replacing $dB_2(t)$ with $i u(t) dt + dB_2(t)$ (displacement, \cite[sec. 9.2.4]{GZ00}). 
Provided the corresponding output channel is not used for further interconnection, this may equivalently be represented by replacing the Hamiltonian term $H=\frac{1}{2} \omega \sigma_z$ in  equation (\ref{eq:hybrid-2}) by the control-dependent Hamiltonian
\begin{equation}
 H(u) = \frac{1}{2} ( \omega \sigma_z + u \sigma_x).
\label{eq:hybrid-4}
\end{equation}
Thus we arrive at the following equations for the controlled atomic system:
\begin{eqnarray}
d U(t) &=& \{ \sqrt{\kappa_1}\, dB_1^\ast(t) \sigma_-  - \sqrt{\kappa_1}\, \sigma_+ dB_1(t) 
+ \sqrt{\kappa_2}\, dB_2^\ast(t) \sigma_-  - \sqrt{\kappa_2}\, \sigma_+ dB_2(t) 
\nonumber \\ 
&& \hspace{2.0cm} 
-  (\frac{1}{2} (\kappa_1+\kappa_2) \sigma_+ \sigma_-   + i H ( u(t) )   ) dt \} U(t),
\ \ \tau_k < t \leq \tau_{k+1},
\nonumber \\
U(\tau_k^+) & = & V_{k} U(\tau_k) .
\label{eq:hybrid-5}
\end{eqnarray}
 The solution $U(t)$ will be determined by a classical control signal $u(\cdot)$ and time-impulse sequence $\gamma$. The controlled two-level atom is illustrated in Figure \ref{fig:atom-1}. Choice of  $u$ and $\gamma$ before being applied to the system (i.e. off-line) is called {\em open loop control}; no feedback of information from the system is used.

What does feedback mean in the context of our two-level  atomic system? The answer depends on how information is extracted from the system and how this information is used to change the behavior of the system. Accordingly, it is helpful to identify the following types of quantum feedback:
\begin{enumerate}
\item
{\em Measurement feedback}. The output field channel, say  $B_{out,1}(t)$,  is continuously monitored (measured) and the measurement  signal is processed by a classical system, called a {\em classical controller} (implemented, say, in classical analog or digital electronics), to determine the closed loop control signals  $u$ and $\gamma$. This type of feedback involves a directional exchange of information between a quantum system and a classical system, and so involves a loss of quantum information. Measurement feedback  is illustrated in Figure \ref{fig:atom-2}.

\begin{figure}[h]
\begin{center}
\includegraphics[scale=0.75]{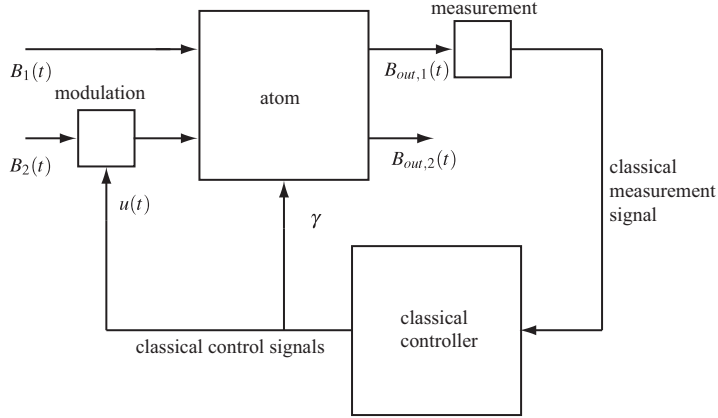}
\caption{Measurement feedback control of a  two-level atom showing measurement of the output channel $B_{out,1}(t)$ field, the classical controller, as well as classical measurement and control signals.}
\label{fig:atom-2}
\end{center}
\end{figure}

\item
{\em Coherent  feedback using quantum signals}. An  output field, say  $B_{out,1}(t)$,  is   not measured, but rather is provided as an input field to another open quantum systems, which we may call a {\em coherent quantum feedback controller}. This quantum controller coherently \lq\lq{processes}\rq\rq \ $B_{out,1}(t)$ 
to produce a field that is shone as an input into the second channel of the atom.
In the coherent feedback loop, information remains at the quantum level, and flows in one direction around the loop.
An example of a coherent feedback arrangement is shown in Figure \ref{fig:atom-3}.

\begin{figure}[h]
\begin{center}
\includegraphics[scale=0.75]{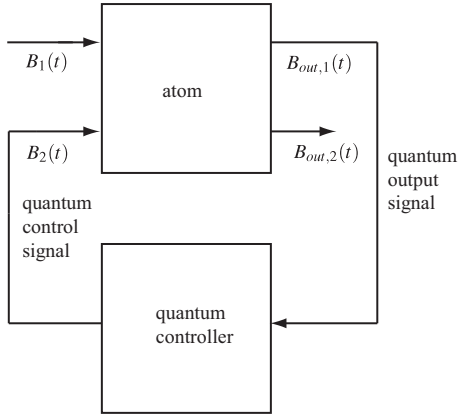}
\caption{Coherent feedback control of a  two-level atom showing  the quantum controller and the quantum signals transmitted around the loop by the fields.  No measurements are  involved.}
\label{fig:atom-3}
\end{center}
\end{figure}

\item
{\em Coherent feedback using direct  coupling}. Here, the atom is simply coupled directly to  another quantum system (without the aid of the fields). This may be regarded as a form of feedback which is bidirectional,  in the spirit of \lq\lq{control by interconnection}\rq\rq \ \cite{JW97}, and does not involve directional quantum signals transmitted via fields.   The second quantum system also serves as a coherent quantum feedback controller, and as with coherent feedback using quantum signals,  all information remains at the quantum level. Direct interaction between the atom and a coherent controller is shown schematically in Figure \ref{fig:atom-4}.

\begin{figure}[h]
\begin{center}
\includegraphics[scale=0.75]{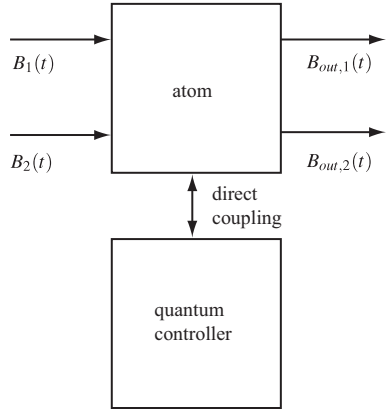}
\caption{Direct coupling of two quantum systems provides a form of feedback control. No signals, and no measurements, are used.}
\label{fig:atom-4}
\end{center}
\end{figure}

\end{enumerate}

There is a large literature on the topic of {\em coherent control}, mainly arising from applications in chemistry and NMR spectroscopy (for example,  \cite{RVMK00}, \cite{KBG01}).
How does this fit into the terminology for control discussed above?  Quite simply: direct couplings between systems may be engaged via the application of a pulse (the coupling is active while the pulse is one, and not active while it is off). The pulse may be regarded as an open loop signal applied to a composite system, resulting in a unitary transformation. In the limit that the pulse has zero width and infinite height, one obtains an impulsive representation of the open loop signal and unitary action.
Note that the quantum controllers discussed above could also depend on classical control signals, although we have not included this possibility explicitly in the above discussion.  So the term \lq\lq{coherent control}\rq\rq \ is rather broadly used, and is meant to convey that the control actions preserve quantum coherence in some way. 

Of course, the two-level model given above may be generalized in a number of ways. For instance,  to cover other physical situations the Hamiltonian $H$ and coupling operators $L_1$ and $L_2$  may be redefined, and the fields may be placed in non-vacuum states.

\section{Open loop control}
\label{sec:open-loop}

Two of the very earliest papers on quantum control are  \cite{VPB79} and  \cite{HTC83}. The paper \cite{VPB79} discusses open quantum models, filtering, and optimal feedback control. This work was well ahead of its time, and was largely unknown for a considerable period. As we will see later on, Belavkin's far-sighted  ideas are highly relevant today. The paper \cite{HTC83}, also very much ahead of its time,  looks at   open loop control of quantum systems, the subject of this section.

\subsection{Bilinear Systems}
\label{sec:open-loop-bilinear}

The simplest type of model for open loop control is that of an isolated two-level atom (no field couplings) with a Hamiltonian $H(u)$ depending on a classical control variable $u$. Indeed, we have the following differential equation for the unitary $U(t)$:
\begin{equation}
\dot{U}(t) = -i \frac{1}{2} (  \omega \sigma_z  + u  \sigma_x   ) U(t), \ \ U(0)=I.
\label{eq:open-1}
\end{equation}

The open loop control system (\ref{eq:open-1}) defines a {\em bilinear} control system evolving on the Lie group $SU(2)$. 
Bilinearity refers to the presence of  products involving the control variables and the variable being solved for (the unitary $U(t)$). 
It is completely deterministic, and beginning with the pioneering paper \cite{HTC83}, a large literature has accumulated studying this type of system using methods from nonlinear control theory, and applying the results  to a range of problems  (see, for example, \cite{DD07}).

Let's now take a closer look at the  dynamics of the atom, first in the {\em Heisenberg picture}, where atomic operators $X$ evolve according to $\dot X = -i [X, H ]$. Since all atomic operators can be expressed in terms of the Pauli matrices, it suffices to determine the dynamics of $X=\sigma_x, \sigma_y$ and $\sigma_z$. By using the commutation relations
\begin{equation}
[ \sigma_x, \sigma_y ] = 2i \sigma_z, \ \ 
[ \sigma_y, \sigma_z ] = 2i \sigma_x, \ \
[ \sigma_z, \sigma_x ] = 2i \sigma_y ,
\label{eq:pauli-commutations}
\end{equation}  
we find that
\begin{equation}
\left(  \begin{array}{c}
\dot{\sigma}_x (t)  \\ \dot{\sigma}_y (t) \\ \dot{\sigma}_z(t)
\end{array} \right) =
\left(  \begin{array}{ccc}
 0 & -\omega & 0
 \\
 \omega & 0 & -u
 \\
 0& u & 0
\end{array} \right) 
\left(  \begin{array}{c}
\sigma_x (t) \\ \sigma_y (t) \\ \sigma_z(t)
\end{array} \right) .
\label{eq:open-2}
\end{equation}
Equation (\ref{eq:open-2})  gives a complete description of the controlled atomic motion, expressed as a bilinear control system in the $\ast$-algebra $\mathscr{M}_2$ (the vector space of $2\times 2$ complex matrices, with the usual matrix multiplication and involution given by the matrix adjoint).

Now we switch to the {\em Schrodinger picture}, within which state vectors evolve as 
$\vert \psi(t) \rangle = U(t) \vert \psi \rangle$, or
$\dot \vert \psi \rangle = -i H \vert \psi \rangle$;  more generally, density operators $\rho$ evolve as $\rho(t)= U(t) \rho U^\ast(t)$, or
$\dot \rho = i[\rho, H]$. Now any density operator may be expressed in the form
\begin{equation}
\rho = \frac{1}{2} ( I + x \sigma_x + y \sigma_y + z \sigma_z ),
\label{eq:open-3}
\end{equation}
where the (real) vector $r=(x,y,z)^T$ is known as the {\em Bloch vector}, with length $\sqrt{x^2 + y^2 +z^2} \leq 1$. The Schrodinger dynamics is given in terms of the Bloch vector as follows:
\begin{equation}
\left(  \begin{array}{c}
\dot{x}(t)   \\ \dot{y} (t) \\ \dot{z}(t)
\end{array} \right) =
\left(  \begin{array}{ccc}
 0 & -\omega & 0
 \\
 \omega & 0 & -u
 \\
 0 & u & 0
\end{array} \right) 
\left(  \begin{array}{c}
x(t) \\ y(t)\\ z(t)
\end{array} \right) .
\label{eq:open-4}
\end{equation}
This equation is a bilinear system in the solid  {\em Block sphere}  $x^2 + y^2 +z^2 \leq 1$.  If the initial Bloch vector is on the  surface of the  Bloch sphere, i.e. $x^2(0) + y^2(0) +z^2(0) = 1$, as is the case for initial state vectors $\vert \psi(0) \rangle$ (i.e. $\rho(0) = \vert \psi(0) \rangle \langle \psi(0) \vert$), then equation (\ref{eq:open-4}) ensures that $x^2(t) + y^2(t) +z^2(t) = 1$ for all $t$, and so describes a bilinear system on the surface of the Bloch sphere.

We see therefore that open loop control of isolated systems leads to interesting bilinear control systems defined on spaces other than Euclidean spaces (Lie groups, $\ast$-algebras, and  manifolds). However, it is important to point out that for some types of quantum systems {\em linear} control systems arise, even though the equations  for the unitary are bilinear. This happens in the case of the quantum harmonic oscillator, where the commutation relation $[a,a^\ast]=1$ (a constant, unlike the commutation relations (\ref{eq:pauli-commutations}) for the Pauli matrices)  gives rise to a linear equation for the annihilation $a$ and creation $a^\ast$  operators.

\subsection{Optimal Control}
\label{sec:open-loop-optimal}

We turn now to the problem of {\em optimally} controlling the atomic system discussed in Section \ref{sec:open-loop-bilinear}. By this we mean (for the moment) to find an open loop control signal $t \mapsto u(t)$  that optimizes a performance criterion chosen to reflect a desired objective. For instance, the paper  \cite{KBG01} used optimal control theory to design pulse sequences to achieve rapid state transfers. The performance criterion used was the time taken to go from an initial unitary to a target final unitary; a problem of {\em time-optimal} control. The authors were able to exploit the rich structure of Lie groups and Lie algebras to develop an elegant formulation of the problem and explicit solutions in some cases. 

It is worth remarking at this point that controllability is closely related to optimal control; indeed, the fundamental ideas of controllability and observability were developed by Kalman in his studies of linear quadratic optimal control problems \cite{REK60a}.  In time-optimal control, the minimum time function $T(x_0)$ is finite precisely when it is possible to steer a system from the initial state to a given target state $x_f$  \cite{EJ89}.

Let's now look at time-optimal control of the atomic system, not at the level of the unitary, but at the level of state vectors. Given a fixed target state $\vert \psi_f \rangle$, and an initial state $\vert \psi_0 \rangle$, find a control signal $u(\cdot)$ that steers the atom from $\vert \psi_0 \rangle$ to $\vert \psi_f \rangle$ in minimum time. In terms of Bloch vectors, given a fixed target state $r_f = (x_f, y_f, z_f)^T$ and an initial state $r_0 = (x_0, y_0, z_0)^T$, find a control signal $u(\cdot)$ that steers the atom (via the dynamics (\ref{eq:open-4})) from $r_0$ to $r_f$ in minimum time. 

In order to formalize this, we define the {\em minimum time function} $T(r)$ (the value function for time-optimal control) by
\begin{equation}
T(r) = \inf_{u(\cdot)} \{ t_f \ : r(0)=r, \ r(t_f) = r_f \}.
\label{eq:min-t-1}
\end{equation}
Here, $t_f$ is the time taken for the atom to move from the initial Bloch vector $r$ to the final Bloch vector $r_f$ using the  control signal $u(\cdot)$. Thus, $T(r)$ is the minimum time over all control signals. 

If there is no restriction on the range of the control signal $u$, then the time-optimal control problem is singular and leads to impulsive solutions,  as in \cite{KBG01}, \cite{SJ08}. A reformulation of this problem using a hybrid model is described in Section \ref{sec:open-loop-impulsive}. For the remainder of this section, let's assume that the controls $u$ take values in  closed interval $\mathbf{U}=[-1,1]$.  
Before proceeding, let's re-write (\ref{eq:open-4}) in compact form
\begin{equation}
\dot{r}(t) = f(r(t), u(t)),
\label{eq:open-5}
\end{equation}
where the vector field $f(r,u)$ is defined by the right hand side of (\ref{eq:open-4}).

{\em Dynamic programming} is a basic tool in optimal control theory, \cite{FR75}, \cite{FS06}. Suppose we have a smooth non-negative solution $S(r)$  to the {\em dynamic programming equation} (DPE) or {\em Hamilton-Jacobi-Bellman}  (HJB)  equation
\begin{eqnarray}
\min_{u \in \mathbf{U}} \{ DS(r) [ f(r,u)] +1 \} &=& 0, 
\label{eq:mintime-2-a}
\\
S(r_f) & = & 0.
\label{eq:mintime-2-b}
\end{eqnarray}
Equation (\ref{eq:mintime-2-a}) is a nonlinear partial differential equation on the Bloch sphere, in which $DS(r)[f(r,u)]$ denotes the directional derivative of the function $S$ at the Bloch  vector $r$ in the direction  $f(r,u)$. Equation (\ref{eq:mintime-2-b})  is a boundary condition, corresponding to the fact that the optimal time taken to go from $r_f$ to itself is zero. 

The main purpose of the DPE is the {\em verification theorem} \cite{FR75}, which allows us to test a candidate optimal control signal for optimality. Indeed, suppose we have a control signal $u^\star(\cdot)$ that attains the minimum in the DPE (\ref{eq:mintime-2-a}), i.e.
\begin{equation}
u^\star(t)  = - \mathrm{sign}( S_z(r^\star(t)) y^\star(t) - S_y(r^\star(t)) z^\star(t) ),
\label{eq:mintime-3}
\end{equation}
where\footnote{sign$(\xi) = +1$ if $\xi \geq 0$ and sign$(\xi)=-1$ if $\xi < 0$.}
 $r^\star(\cdot)$ is the corresponding trajectory of (\ref{eq:open-5}) with initial condition $r^\star(0)=r$. Then $u^\star(\cdot)$ is optimal and $S(r)$ equals $T(r)$, the minimum time function defined by (\ref{eq:min-t-1}). 

To see why the verification theorem is true, let $u(\cdot)$ be any control signal steering $r$ to $r_f$, and let $0 < t < t_f$. Now integrate equation (\ref{eq:mintime-2-a})  along the trajectory to obtain
\begin{eqnarray}
S(r(t)) &=&  S(r) + \int_0^t  DS(r(s))[f(r(s), u(s))] ds
\nonumber \\
& \geq & S(r) - t
\label{eq:mintime-4}
\end{eqnarray}
with equality if $u(\cdot)=u^\star(\cdot)$. Setting $t=t_f$ we see that $S(r)\leq t_f$ with equality if $u(\cdot)=u^\star(\cdot)$. Hence $S(r)=T(r)$ and $u^\star(\cdot)$ is optimal.

Expression (\ref{eq:mintime-3}) suggests  a formula for \lq\lq{feedback}\rq\rq \  optimal controls. Define
\begin{equation}
\mathbf{u}^\star(r)  = - \mathrm{sign} ( S_z(r)y - S_y(r)  z),
\label{eq:mintim-5}
\end{equation}
that is, for any Bloch vector $r=(x,y,z)^T$, $\mathbf{u}^\star(r)$ is a control value that attains the minimum of $DS(r)[ f(r,u) ]$. However, this \lq\lq{feedback}\rq\rq \ formula requires knowledge of the Bloch vector $r$, which is not possible in the present context - no measurement information is available, and the quantum state is not a measurable quantity. However, expression (\ref{eq:mintime-3}) may be used off-line, in a computer simulation to determine an optimal {\em open loop} control signal. If one wishes, the optimizing control may be substituted into the DPE (\ref{eq:mintime-2-a})  to re-write it in the form
\begin{equation}
S_x(r) (-\omega y) + S_y(r)(\omega x) - \vert S_z(r) y - S_y(r) z \vert +1 =0.
\label{eq:mintime-5a}
\end{equation}



Where does the DPE (\ref{eq:mintime-2-a}) come from? Well, if the minimum time function is sufficiently smooth, then it solves the DPE (\ref{eq:mintime-2-a}) (by definition it satisfies (\ref{eq:mintime-2-b})). To see this, for any $t>0$ the minimum time function satisfies
\begin{equation}
T(r) = \min_{u(\cdot)} \{ \min(t,t_f)  +T(r(\min(t,t_f)) ) \ : r(0)=r \} .
\label{eq:mintim-5a}
\end{equation}
Then if $t < t_f$ we may differentiate (\ref{eq:mintim-5})  to obtain (\ref{eq:mintime-2-a}).

An important technical issue is that in general $T$ is not everywhere differentiable, and nonlinear PDEs like (\ref{eq:mintime-2-a}) do not in general admit smooth solutions. Nevertheless, $T$ solves (\ref{eq:mintime-2-a}) in a weaker sense  that does not require smoothness. The theory of viscosity solutions was developed to deal with nonsmooth solutions to nonlinear PDE, \cite{FS06}.

In order to gain some more insight into the nature of the solution to the optimal control problem, we switch to polar coordinates and use the fact that the dynamics  (\ref{eq:open-4}) preserves states on the surface of the Bloch sphere. We write $x=\sin \theta \cos \phi$, $y=\sin \theta \sin \phi$, $z=\cos \theta$, and find that the dynamics becomes
\begin{eqnarray}
\dot \theta(t) &=& - u(t) \sin \phi(t)
\label{eq:mintim-polar-1-theta}
\\
\dot \phi(t) &=& - u(t) \cot \theta(t) \cos \phi(t) + \omega.
\label{eq:mintim-polar-1-phi}
\end{eqnarray}
In these polar coordinates, the target state is $(\theta_f, \phi_f)$, and if
we write $\tilde S(\theta,\phi) = S( \sin \theta \cos \phi, \sin \theta \sin \phi, \cos \theta)$ we obtain  the DPE
\begin{eqnarray}
\min_{u \in \mathbf{U}} \{ -u ( \tilde S_\theta(\theta, \phi)  \sin \phi   +\tilde S_\phi (\theta,\phi)   \cot \theta \cos \phi ) + \tilde S_\phi(\theta, \phi) \omega
+1 \} &=& 0, 
\label{eq:mintime-polar-2-a}
\\
\tilde S(\theta_f, \phi_f) & = & 0.
\label{eq:mintime-polar-2-b}
\end{eqnarray}
The control attaining the minimum in (\ref{eq:mintime-polar-2-a}) is 
\begin{equation}
\mathbf{u}^\star (\theta, \phi) = \mathrm{sign} (  \tilde S_\theta(\theta, \phi)  \sin \phi   +\tilde S_\phi (\theta,\phi)   \cot \theta \cos \phi ).
\label{eq:mintime-polar-3}
\end{equation}



\subsection{Impulsive Control}
\label{sec:open-loop-impulsive}

Suppose we remove the restriction on the range of the control signal, i.e. take $\mathbf{U}=\mathbb{R}$. Now the vector field $f(r,u)$ has the form $f(r,u)=f_0(r) + f_1(r)u$, and so if we attempted to find the minimum in the DPE (\ref{eq:mintime-2-a}) we would find that it is not defined, as the controls would need to be infinitely large. This singular situation leads us to impulsive control actions, \cite{KBG01}, \cite{SJ08}, which form the subgroup $\mathbf{V}$ of $SU(2)$ determined by the control Hamiltonian $\sigma_x u$, that is,
\begin{equation}
\mathbf{V}= \{ e^{-i v  \sigma_x} \ : \  v  \in \mathbb{R} \}.
\label{eq:impulse-0}
\end{equation}

Now let's  use a hybrid continuous-impulsive model for time-optimal control on the surface of the Bloch sphere. Let $\gamma$ be an impulsive open loop control of the form $\gamma = ( (\tau_0, v_0), (\tau_1, v_1), \ldots ) $ (as in equation (\ref{eq:hybrid-3})), and consider the hybrid form of the Schrodinger equation
\begin{eqnarray}
\dot{U}(t) &=& -i \frac{1}{2} \omega \sigma_z \, U(t), \ \ \tau_k < t \leq \tau_{k+1},
\nonumber
\\
U(\tau^+_k) &=& e^{-i v_k \sigma_x } U(\tau_k),
\label{eq:impulse-1}
\end{eqnarray}
 where $V_k(\tau_k) = U^\ast(\tau_k) V_k U(\tau_k) = e^{-i v_k \sigma_x(\tau_k)}$  is the impulse applied  at time $\tau_k$.

The hybrid equations of motion on the surface of the  Block sphere are 
\begin{eqnarray}
\left(  \begin{array}{c}
\dot{x}(t)   \\ \dot{y} (t) \\ \dot{z}(t)
\end{array} \right) &=& 
\left(  \begin{array}{ccc}
 0 & -\omega & 0
 \\
 \omega & 0 & 0
 \\
0 & 0 & 0
\end{array} \right) 
\left(  \begin{array}{c}
x(t) \\ y(t)\\ z(t)
\end{array} \right),    \ \ \tau_k < t \leq \tau_{k+1},
\label{eq:impulse-2-a} \\
\left(  \begin{array}{c}
x(\tau^+_k ) \\ y( \tau^+_k )\\ z(\tau^+_k)
\end{array} \right) 
&=&
\left(  \begin{array}{ccc}
 1 & 0 & 0
 \\
0 & \cos v_k  & -\sin v_k
 \\
 0   & \sin v_k & \cos v_k
\end{array} \right) 
\left(  \begin{array}{c}
x(\tau_k ) \\ y( \tau_k )\\ z(\tau_k)
\end{array} \right).
\label{eq:impulse-2-b}
\end{eqnarray}
These equations describe the natural drift, a rotation in the $xy$ plane, together with a choice of instantaneous rotation in the $yz$ plane produced by the selected impulse. 

The minimum time function $T(r)$ is again defined by (\ref{eq:min-t-1}), but now the DPE takes the form
\begin{eqnarray}
\min \{ DS(r) [ f_0 (r) ] +1, \inf_{V \in \mathbf{V}}   S(VU)-S(U)   \} &=& 0, 
\label{eq:impulse-3-a}
\\
S(r_f) & = & 0,
\label{eq:impulse-3-b}
\end{eqnarray}
to which one seeks a non-negative solution. If the minimum time function $T(r)$  is sufficiently smooth, then it will be a solution of the {\em quasivariational inequality}  (QVI) (\ref{eq:impulse-3-a}). The QVI has two parts:
\begin{eqnarray}
  DS(r) [ f_0 (r) ] +1 \geq 0 \ & \mathrm{and}& \  \min_{V \in \mathbf{V}}   S(VU) \geq S(U)    ,
\label{eq:impulse-4-a}
\\
   DS(r) [ f_0 (r) ] +1 = 0 \ & \mathrm{or}& \  \min_{V \in \mathbf{V} }   S(VU) =  S(U) .
\label{eq:impulse-4-b}
\end{eqnarray}
Equation (\ref{eq:impulse-4-a}) simply says that drifting or impulsing will lead to a time greater than or equal to the minimum time. Equation  (\ref{eq:impulse-4-b}) says that along the optimal trajectory the atom should drift at Block vectors  for which $DS(r) [ f_0 (r) ] +1 = 0$, while if the Bloch vector is such that $\min_{V \in \mathbf{V}}   S(VU) =  S(U)$, then the impulse $V \in \mathbf{V}$ should be applied.


\subsection{Relaxation}
\label{sec:open-loop-relax}

Let's  take a look at open loop control of the atom (or ensemble of atoms) in the presence of a decohering mechanism.  For definiteness,  consider an atom coupled to a single field channel  ($\kappa=\kappa_1$, $B(t)=B_1(t)$ in the notation of Section \ref{sec:types}), with impulsive control only (with impulses in $\mathbf{V}$, the subgroup defined by (\ref{eq:impulse-0})). The hybrid Schrodinger equation for the unitary is
\begin{eqnarray}
d U(t) &=& \{ \sqrt{\kappa}\, dB^\ast(t) \sigma_-  - \sqrt{\kappa}\, \sigma_+ dB(t) 
- \frac{1}{2}   (\kappa \sigma_+ \sigma_-   + i  \omega \sigma_z )dt  )\} U(t),
\nonumber \\ 
&& \hspace{6.0cm} 
 \tau_k < t \leq \tau_{k+1},
\nonumber \\
U(\tau_k^+) & = & V_{k} U(\tau_k) .
\label{eq:relax-1}
\end{eqnarray}

In the Heisenberg picture, atomic operators $X$ evolve according to $X(t)=U^\ast(t) X U(t)$, so that between impulses we have
\begin{eqnarray}
dX(t) &=&  (-i [ X(t), H(t) ] + \mathcal{L}_{L(t)} (X(t)))dt 
\nonumber \\
&& \ \  + dB^\ast(t) [ X(t), L(t) ] + [ L^\ast(t), X(t)] dB(t), \ \tau_k < t \leq \tau_{k+1},
\label{eq:relax-2}
\end{eqnarray}
where $L=\sqrt{\kappa}\, \sigma_-$, and
\begin{equation}
\mathcal{L}_L(X) = \frac{1}{2} L^\ast [ X, L ] + \frac{1}{2} [ L^\ast, X ] L.
\label{eq:relax-3}
\end{equation}
When an impulse is applied at time $\tau_k$, we have
\begin{equation}
X(\tau_k^+) = U^\ast(\tau_k^+) X U(\tau_k^+) = V_k^\ast (\tau_k) X(\tau_k) V_k(\tau_k),
\label{eq:relax-3a}
\end{equation}
where $V_k(\tau_k) = U^\ast(\tau_k) V_k U(\tau_k) = e^{-i v_k \sigma_x(\tau_k)}$.

 Explicitly, for the operators $X=\sigma_x, \sigma_y$ and $\sigma_z$, we have,  for $\tau_k < t  \leq \tau_{k+1}$,
 \begin{eqnarray}
 d \sigma_x(t) &=&  ( -\omega \sigma_y(t) - \frac{\kappa}{2} \sigma_x(t)) dt + \sqrt{\kappa}  (dB^\ast(t)\sigma_z(t) + \sigma_z(t) dB(t)     )
\label{eq:relax-4-x}  \\
 d \sigma_y(t) &=& (\omega \sigma_x(t) - \frac{\kappa}{2} \sigma_y(t)) dt  -  i \sqrt{\kappa}   ( \sigma_z(t) dB^\ast(t)   - \sigma_z(t) dB(t) )
 \label{eq:relax-4-y}
 \\
 d \sigma_z(t) &=& (-  \kappa \sigma_z(t) -\kappa  ) dt - 2 \sqrt{\kappa} (dB^\ast(t)  \sigma_-(t)  +  \sigma_+ (t) dB(t)),
 \label{eq:relax-4-z}
 \end{eqnarray}
 and for $t=\tau_k^+$,
 \begin{eqnarray}
 \sigma_x( \tau_k^+) &=&  \sigma_x( \tau_k),
   \label{eq:relax-4-impulse-x}
 \\
  \sigma_y( \tau_k^+) &=&   \cos( v_k) \sigma_y( \tau_k) - \sin( v_k) \sigma_z( \tau_k)  ,
    \label{eq:relax-4-impulse-y}
  \\
   \sigma_z( \tau_k^+) &=& \sin( v_k) \sigma_y( \tau_k) + \cos( v_k) \sigma_z( \tau_k)  .
  \label{eq:relax-4-impulse-z}
 \end{eqnarray}
 Equations (\ref{eq:relax-4-x})-(\ref{eq:relax-4-impulse-z}) constitute a set of impulsive QSDEs in the $\ast$-algebra $\mathscr{M}_2 \otimes \mathscr{F}$, where $\mathscr{F}$ is the $\ast$-algebra of field operators (defined on an underlying Fock space).
By taking expectations, we may conclude from (\ref{eq:relax-4-z})  that the mean energy of the atom decreases exponentially, that is, the atom looses energy to the field. 

The equations of motion for the atomic state $\rho$  may be obtained by averaging out the noise in the Schrodinger picture, $\rho(t) = \mathrm{tr}_B[ U(t) (\rho \otimes \vert 0 \rangle \langle 0 \vert ) U^\ast(t) ]$; here, we take the field to be in the vacuum state. The differential equation for $\rho(t)$, which holds between impulses,  is
\begin{equation}
\dot{\rho}(t) = i [ \rho(t), H ] + \mathcal{L}^\ast_{L}( \rho(t)) 
\label{eq:relax-5}
\end{equation}
where
\begin{equation}
\mathcal{L}^\ast_L(\rho) = \frac{1}{2}  [L, \rho L^\ast ] + \frac{1}{2} [ L\rho, L^\ast ].
\label{eq:relax-6}
\end{equation}
Equation (\ref{eq:relax-5}) is called the {\em master equation}, and $i[\rho,H] + \mathcal{L}^\ast_L(\rho)$ is called the {\em Lindblad superoperator} (in Schrodinger form). 

The hybrid equations of motion inside the  Bloch sphere  are
\begin{eqnarray}
\dot{x}(t) &=& - \frac{\kappa}{2} x(t) -\omega y(t) , \ \ \tau_k < t \leq \tau_{k+1}, 
\label{eq:relax-7-a} \\
\dot{y}(t) &=& - \frac{\kappa}{2} y(t) + \omega x(t)  , \ \ \tau_k < t \leq \tau_{k+1},
\label{eq:relax-7-b} \\
\dot{z}(t) &=&  -\kappa z(t) - \kappa, \ \ \tau_k < t \leq \tau_{k+1},
\label{eq:relax-7-c} \\
\left(  \begin{array}{c}
x(\tau^+_k ) \\ y( \tau^+_k )\\ z(\tau^+_k)
\end{array} \right) 
&=&
\left(  \begin{array}{ccc}
 1 & 0 & 0
 \\
0 & \cos v_k  & -\sin v_k
 \\
 0   & \sin v_k & \cos v_k
\end{array} \right) 
\left(  \begin{array}{c}
x(\tau_k ) \\ y( \tau_k )\\ z(\tau_k)
\end{array} \right).
\label{eq:relax-7-d}
\end{eqnarray}
In equations (\ref{eq:relax-7-a})-(\ref{eq:relax-7-c}) we can see the effect of the field coupling, which in the absence of control action causes $x(t) \to 0$, $y(t)\to 0$ and $z(t)\to -1$ as $t \to \infty$, i.e. $\rho(t)\to \frac{1}{2}( \sigma_z - I)= \vert -1 \rangle \langle -1 \vert$, the pure state of lowest energy. Repeated application of impulses offers the possibility of achieving other large time behaviour.
For instance, the periodic pulse sequence $\gamma$ with period $1$ and $v_k=\pi/2$ leads to the steady state $(0, 0.5168, 0.3135)$ (for the case $\kappa=1$, $\omega=0$).


\section{Quantum  Feedback Networks}
\label{sec:networks}

A glance at the various types of feedback control discussed in 
Section \ref{sec:types}, or indeed any textbook on classical  feedback control,  tells us that feedback arrangements are {\em networks} of interconnected systems. Accordingly, the purpose of this section is to set up some easy-to-use  tools for constructing feedback networks. What is presented in this section is a simplification of a more general quantum feedback network theory \cite{YK03a},  \cite{GJ09},  \cite{GJN09a}, which builds on earlier cascade theory \cite{CWG93}, \cite{HJC93}, network quantization \cite{YD84}, and quantum control \cite{WM94b}.

\begin{figure}[h]
\begin{center}
\includegraphics{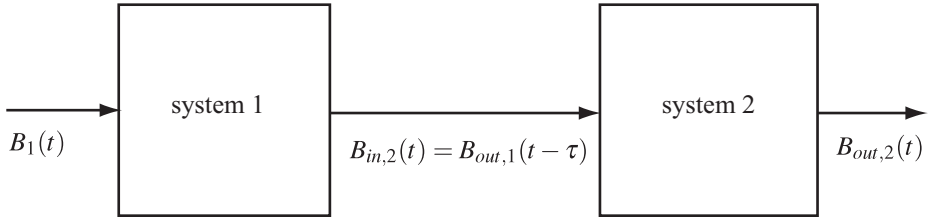}
\caption{A series connection between two systems.}
\label{fig:qfn-1}
\end{center}
\end{figure}

The basic idea is simple, Figure \ref{fig:qfn-1}. Take an output channel and connect it to an  input channel. Such series or cascade connections are commonplace in classical electrical circuit theory. For instance, if the systems are resistors with resistances $R_1$ and $R_2$, then the total series-connected system is equivalent to a single resistor with resistance $R=R_1+R_2$. This use of simple parameters for devices, and rules for interconnecting devices in terms of these parameters, is a powerful feature of classical electrical circuit theory.

Let's see how we can achieve an analogous simple rule for the series connection of two open quantum systems (without impulses), 
where the input and output channels are quantum fields, as discussed in Section \ref{sec:types}. The physical parameters 
determining open quantum systems are a Hamiltonian $H$ describing the self-energy of the system, and a vector $L$ of operators describing how the system is coupled to the field channels. These parameters appear in the Lindblad superoperator $i[\rho,H] + \mathcal{L}^\ast_L(\rho)$ (and hence in the master equation), as well as in QSDEs, \cite{HP84}, \cite{GC85}, \cite{KRP92}, \cite{GZ00}.  
Actually, there is a third parameter $S$, a self-adjoint matrix of operators describing scattering between field channels, that was introduced in \cite{HP84},  \cite{KRP92}. While $S$ does not appear in the Linblad nor the master equation for a single system, it does have non-trivial use in several applications   including quantum feedback networks (such as those including beamsplitters).  Thus in general an open quantum system, call it $G$, is characterized by three parameters $G=(S,L,H)$. However, in this article we do not use the scattering parameter, and so we set $S=I$; actually, we make the abbreviation $G=(L,H)$.
For example, the parameters for the atom,  coupled to two field channels  (recall section \ref{sec:types}), call it $A$, are
\begin{equation}
A = \left(   
\left( \begin{array}{c}
\sqrt{\kappa_1}\, \sigma_-
\\
\sqrt{\kappa_2}\, \sigma_-
\end{array} \right),
\frac{1}{2}\omega \sigma_z
\right).
\label{eq:qfn-1}
\end{equation}

In network modeling, and indeed in modeling in general, it can be helpful to decompose large systems into smaller pieces, and to assemble large systems from components. In \cite{GJ09}, the {\em concatenation product} $\boxplus$ was introduced  to assist with this. The concatenation product is defined by
\begin{equation}
(L_1, H_1)  \boxplus (L_2, H_2) = \left( 
\left( \begin{array}{c}
L_1
\\
L_2
\end{array} \right),
H_1+H_2 
\right).
\label{eq:qfn-2}
\end{equation}
For the atom, if we wish to decompose it with respect to field channels we may write
\begin{equation}
A = ( \sqrt{\kappa_1}\, \sigma_-, \, \frac{1}{2}\omega \sigma_z ) \boxplus ( \sqrt{\kappa_2}\, \sigma_-, 0) .
\label{eq:qfn-3}
\end{equation}

Now suppose we have two systems $G_1 = (L_1, H_1)$ and $G_2 = (L_2, H_2)$, as in Figure \ref{fig:qfn-1}. Because the systems are separated spatially, the field segment connecting connecting the output of $G_1$ to the input of $G_2$ has non-zero length, and so this means there is a small delay $\tau$ in the transmission of quantum information from $G_1$ to $G_2$. That is, $B_{in,2}(t) = B_{out,1}(t-\tau)$. Now if the systems are sufficiently close, $\tau$ will be small compared with the timescales of the systems, and may be neglected. In this way, a Markovian model for the series connection $G= G_2 \triangleleft G_1$ may be derived. In terms of the physical parameters, the {\em series product} (defined in  \cite{GJ09}) is given by
\begin{equation}
(L_2, H_2)  \triangleleft  (L_1, H_1) = 
(L_1 + L_2, H_1+H_2 + \mathrm{Im}[ L_2^\dagger L_1]).
\label{eq:qfn-4}
\end{equation}
Thus the series connection $G= G_2 \triangleleft G_1$ has parameters $L=L_1+L_2$ and $H=H_1+H_2 + \mathrm{Im}[ L_2^\dagger L_1]$, analogous to the expression $R=R_1+R_2$ for series-connected resistors.

The series product serves very well for Markovian approximations to cascades of independent open systems. However, importantly for us, the series product may also be used to describe an important class of {\em quantum feedback networks}. This is because the two systems 
$G_1 = (L_1, H_1)$ and $G_2 = (L_2, H_2)$ need not be independent---they can be parts of the same system. 

\begin{figure}[h]
\begin{center}
\includegraphics{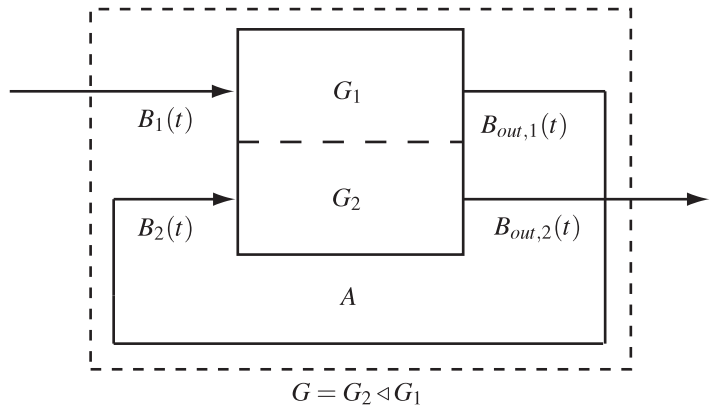}
\caption{Atom in a coherent feedback loop described by a series connection $G=G_2 \triangleleft G_1$.}
\label{fig:qfn-2}
\end{center}
\end{figure}

For example, suppose we take $G_1$ to be the first factor in the decomposition (\ref{eq:qfn-3})  of $A$, i.e. $G_1=( \sqrt{\kappa_1}\, \sigma_-, \, \frac{1}{2}\omega \sigma_z)$, and $G_2= ( \sqrt{\kappa_2}\, \sigma_-, 0)$ the second. Then the series connection 
\begin{equation}
G= ( \sqrt{\kappa_2}\, \sigma_-, 0)   \triangleleft  ( \sqrt{\kappa_1}\, \sigma_-, \, \frac{1}{2}\omega \sigma_z) = 
\left( (\sqrt{\kappa_1} + \sqrt{\kappa_2}) \sigma_-, \, \frac{1}{2}\omega \sigma_z
\right)
\label{eq:qfn-5}
\end{equation}
describes the coherent feedback arrangement shown in Figure \ref{fig:qfn-2}. The system $G$ is an open quantum system with Hamiltonian $H=\frac{1}{2}\omega \sigma_z$ that is coupled to a single field channel via the coupling operator $L = (\sqrt{\kappa_1} + \sqrt{\kappa_2}) \sigma_-$.
Now that we have the parameters for the feedback system $G$, it is easy to write down the corresponding Schrodinger equation
\begin{eqnarray}
dU(t) &=&  \{   
 (\sqrt{\kappa_1} + \sqrt{\kappa_2}) ( dB^\ast(t) \sigma_- - \sigma_+ dB(t) )
 \nonumber \\
 && \hspace{2.0cm}  - \frac{1}{2} (
  (\sqrt{\kappa_1} + \sqrt{\kappa_2})^2 \sigma_+ \sigma_- + i \omega \sigma_z )dt
 \} U(t)
\label{eq:qfn-6}
\end{eqnarray}
and   master equation
\begin{equation}
\dot{\rho} = i[ \rho, \frac{1}{2}\omega \sigma_z ] + \frac{1}{2}(\sqrt{\kappa_1} + \sqrt{\kappa_2})^2 (  [ \sigma_-, \rho \sigma_+]
+ [ \sigma_- \rho, \sigma_+]),
\label{eq:qfn-7}
\end{equation}
if desired. Equations like these apply to complete systems, and have meaning only when the network construction process has  concluded.

To further illustrate the use of the series product, consider a situation with $L_1 = \sqrt{\kappa_1}\, \sigma_-$ and
  $L_2=i\sqrt{\kappa_2}\, \sigma_z$.
  Then the Markovian model for the series connection
  \begin{equation}
G = ( i\sqrt{\kappa_2}\, \sigma_z, 0)  \triangleleft  (\sqrt{\kappa_1}\, \sigma_-, \frac{1}{2} \omega \sigma_z) =
( (\sqrt{\kappa_1}\, \sigma_- +  i\sqrt{\kappa_2}\, \sigma_z, \frac{1}{2} \omega \sigma_z + \frac{1}{2} \sqrt{\kappa_1+\kappa_2}\,  \sigma_x)
\label{eq:qfn-8}
\end{equation}
contains an additional Hamiltonian term $\frac{1}{2} \sqrt{\kappa_1+\kappa_2}\,  \sigma_x$. The reader may wish to derive the corresponding Schrodinger and master equations from these parameters.

\section{Quantum Filtering}
\label{sec:filter}

The term {\em filtering} is used in many ways, but usually refers to a process of extracting information concerning something of interest from a  source containing partial information that may be noisy. Our interest here is in the extraction of classical information about the atom   by monitoring the output field channel $B_{out,1}(t)$. For instance, we may wish to know something about the atom's energy by observing any photons emitted into the field.
The {\em quantum filter} was developed for purposes like this \cite{VPB92a}, and also  goes by the name {\em stochastic master equation} \cite{WM10,HC93}.

\subsection{Probability}
\label{sec:filter-probability}

In quantum mechanics the postulates state that physical quantities are represented by {\em observables}, which are self-adjoint operators defined on some underlying Hilbert space.
Consider for a moment an isolated atom.
Atomic energy is represented by the observable $\frac{1}{2}\omega \sigma_z$.  The possible measurement outcomes are the eigenvalues of the observable, which are $\pm \frac{1}{2}\omega$ for the energy of the atom. The probabilities of the measurement outcomes depends on the state $\rho$, and are given by $\mathrm{Prob}( \pm  \frac{1}{2}\omega) = \mathrm{tr}[ \rho P_{\pm 1}]$, where $P_{\pm 1} = \vert \pm 1 \rangle \langle \pm 1 \vert$ are the projection  operators arising in the spectral representation $\sigma_z = P_{+1} - P_{-1}$. If the outcome $\pm \frac{1}{2}\omega$ is observed, then Von Neumann's projection postulate states that the atomic state changes to $\frac{P_{\pm 1} \rho P_{\pm 1}}{\mathrm{tr}[ \rho P_{\pm 1}] }$;  this is called a {\em conditional state}.

The mathematics that underlies the measurement postulate is the {\em spectral theorem}, which says essentially that any collection of commuting matrices can be simultaneously diagonalized. For instance, let $\mathscr{C}$ be a collection of $2 \times 2$ complex matrices \lq\lq{generated}\rq\rq \ by $\sigma_z$, the energy observable. This means that $\mathscr{C}$ contains all complex linear combinations of powers   $\sigma_z$, and all adjoints (mathematically, $\mathscr{C}$ is called a $\ast$-algebra, a vector space closed under products and adjoints).   The spectral theorem says that any matrix $X \in \mathscr{C}$  can be   diagonalized to a matrix of the form $\mathrm{diag}(a,b)$, where $a$ and $b$ are complex numbers; that is, $X=a P_{+1} + b P_{-1}$.

We can interpret the spectral theorem probabilistically, consistent with the measurement postulate,  as follows. 
For any matrix $X \in \mathscr{C}$
we can define a function $\iota(X)$ on a set $\Omega = \{ \pm 1 \}$ by $\iota(X)(+1)=a$, $\iota(X)(-1)=b$. 
The set
$\Omega$ is a sample space, and $\iota(X)$ is a classical random variable.  The algebra $\mathscr{C}$ and the density operator $\rho$ determine a 
 classical probability distribution $\mathbf{P}$:  $\mathrm{Prob}( \iota(X) =a) =\mathbf{P}(+1) = \mathrm{tr}[ \rho P_{+1}]$, and $\mathrm{Prob}( \iota(X)=b) =\mathbf{P}(-1) = \mathrm{tr}[ \rho P_{-1}]$.  
In what follows we denote quantum expectations (states) by
\begin{equation}
\mathbb{P} [X ] = \mathrm{tr}[\rho X],
\label{eq:prob-1}
\end{equation}
where $X$ is an operator, and classical expectations by
\begin{equation}
\mathbf{P} [ X ] = \sum_{\omega \in \Omega} X(\omega) \mathbf{P}  (\omega ) = \int_\Omega X (\omega)  \mathbf{P}(d\omega),
\label{eq:prob-2}
\end{equation}
where $X$ is a classical random variable.
The spectral theorem  may be re-stated as linking quantum and classical expectations: roughly,
for any $X \in \mathscr{C}$ there exists a classical random variable $\iota(X)$  and a classical probability distribution $\mathbf{P}$ such that
\begin{equation}
\mathbb{P} [X ]  = \mathbf{P} [  \iota(X) ] .
\label{eq:prob-3}
\end{equation}
This can be easily seen, since any $X \in \mathscr{C}$  has the diagonal representation $X=a P_{+1} + b P_{-1}$, and hence $\iota(X)= a \chi_{+1} + b \chi_{-1}$, where the indicator functions $\chi_{\pm 1}$ are defined by $\chi_{\omega_0}(\omega)=1$ when $\omega=\omega_0$, and $\chi_{\omega_0}(\omega)= 0$ when $\omega\neq \omega_0$ ($\omega_0 = \pm 1$).
It is important to appreciate  that while the quantum expectation (\ref{eq:prob-1}) is defined for any $2 \times 2$ complex matrix $X \in \mathscr{M}_2$, expression (\ref{eq:prob-3}) depends on the choice of commutative algebra $\mathscr{C}$, or observable, which in turn determines the classical probability.

\begin{figure}[h]
\begin{center}
\includegraphics{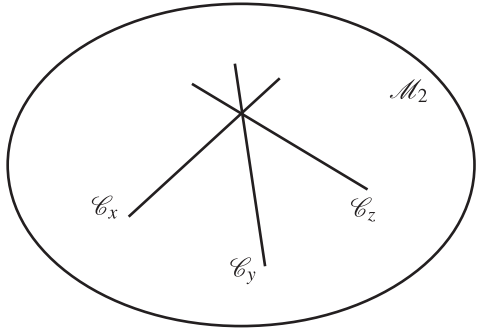}
\caption{The non-commutative quantum probability space $(\mathscr{M}_2, \mathbb{P})$   contains distinct commutative quantum probability spaces $(\mathscr{C}_x, \mathbb{P})$, $(\mathscr{C}_y, \mathbb{P})$ and $(\mathscr{C}_z, \mathbb{P})$, determined by non-commuting observables $\sigma_x$, $\sigma_y$ and $\sigma_z$ respectively. These 
 correspond to distinct classical probability spaces $(\Omega_x, \mathcal{F}_x, \mathbf{P}_x)$, $(\Omega_y, \mathcal{F}_y, \mathbf{P}_y)$ and $(\Omega_z, \mathcal{F}_z, \mathbf{P}_z)$ respectively, which may describe distinct experiments.
}
\label{fig:filter-m2}
\end{center}
\end{figure}

It is sometimes helpful to write $(\mathscr{A}, \mathbb{P})$ for a  {\em quantum probability  space} (where $\mathscr{A}$ is a   $\ast$-algebra (not commutative in general), and $\mathbb{P}$ is quantum expectation), in contrast to a {\em classical probability space} $(\Omega, \mathcal{F}, \mathbf{P})$ (where $\Omega$ is a sample space, $\mathcal{F}$ is a $\sigma$-algebra of events, and $\mathbf{P}$ is a classical probability measure). 
The spectral theorem says that a {\em commutative} quantum probability space $(\mathscr{C}, \mathbb{P})$ is {\em statistically equivalent}  to a classical probability space $(\Omega, \mathcal{F}, \mathbf{P})$. 
In general, a non-commutative quantum probability space can contain many distinct commutative subspaces, each of which is equivalent to distinct classical probability spaces, Figure \ref{fig:filter-m2}.

\subsection{Conditional Expectation}
\label{sec:filter-conditional}
 
  In classical probability information is summarized by $\sigma$-algebras  ${\mathcal G} \subset \mathcal F$ of events in a classical probability space $(\Omega, {\mathcal F}, {\mathbf P})$. 
  Typically, $\mathcal{G}$ will  be the $\sigma$-algebra $\sigma(X)$ generated by a random variable $X$, which contains information on the values taken by $X$. The mathematical notion of {\em measurability} with respect to a $\sigma$-algebra plays an important role in integration theory and probability theory.\footnote{The mathematical term {\em measurable}   is not to be confused 
with measurements in quantum mechanics.}
 If $Z$ is a random variable  measurable with respect to a $\sigma$-algebra $\mathcal{G}=\sigma(X)$ generated by a random variable $X$, then $Z = f(X)$ for some function $f$.

Conditional expectation plays a fundamental role in classical estimation and filtering. Suppose we are given two random variables 
$X$ and $Y$. The random variable $X$ may describe some quantity that is not directly accessible by experiment;  instead, a quantity described by $Y$ is accessible---its values may be obtained by experiment. Then given an outcome $y$ (a value of $Y$), one may wish to improve one's estimation of the expected value of $X$. Now the information associated with $Y$ is described by the $\sigma$-algebra generated by $Y$, $\mathscr{Y}=\sigma(Y)$, and the {\em conditional expectation} of $X$ given $Y$ is the random variable $\mathbf{P}[ X \vert \mathscr{Y} ]$ (often denoted $\mathbf{P}[X \vert Y ]$ or $\hat X$\footnote{Not to be confused with the hats sometimes used in the physics literature to denote operators.}). The conditional expectation $\mathbf{P}[ X \vert \mathscr{Y} ]$ is the unique $\mathscr{Y}$-measurable random variable    such that
\begin{equation}
\mathbf{P}[  \chi_E  \mathbf{P}[ X \vert \mathscr{Y} ] ] = \mathbf{P}[ \chi_E X ]  \ \ \text{for all events} \ E \in \mathscr{Y},
\label{eq:qm-prob-cexp-1}
\end{equation}
where $\chi_E$ is the indicator function for the event $E$ ($\chi_E(\omega) =1$ if $\omega \in E$, zero otherwise).
If $X$ and $Y$ have a joint density $p_{X,Y}(x,y)$, then the conditional density is given by
\begin{equation}
p_{X\vert Y}(x \vert y) = \frac{ p_{X,Y}(x,y)  }{  \int p_{X,Y}(x,y)  dx } ,
\label{eq:qm-prob-cexp-1a}
\end{equation}
 from which the conditional expectation can be computed:
\begin{equation}
\mathbf{P}[ X \vert \mathscr{Y}  ] (y) = \mathbf{P}[ X \vert \, Y=y ] = \int  x p_{X\vert Y}(x \vert y) dx .
\label{eq:qm-prob-cexp-2}
\end{equation}
Equation (\ref{eq:qm-prob-cexp-2}) shows explicitly that the conditional expectation is a function of the outcomes $y$.

\begin{figure}[h]
\begin{center}
\includegraphics{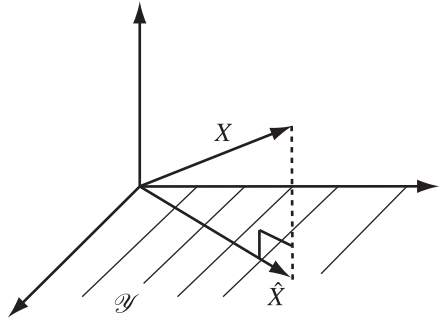}
\caption{The conditional expectation $\hat X = \mathbf{P}[ X \vert \mathscr{Y}  ] $ is the orthogonal projection of $X$ onto the subspace $\mathscr{Y}$.
}
\label{fig:c-exp-1}
\end{center}
\end{figure}

A well known property of the  conditional expectation $\mathbf{P}[ X \vert \mathscr{Y}  ] $ is that it is a {\em minimum variance} or {\em least squares estimator}. Geometrically, $\hat X=\mathbf{P}[ X \vert \mathscr{Y}  ] $ is the orthogonal projection of $X$ onto the subspace $\mathscr{Y}$, Figure \ref{fig:c-exp-1} (see, e.g. \cite{AM79}).

Does conditional expectation make sense in quantum mechanics? The answer is yes, provided we don't try to project all operators at the same time. Given a commutative $\ast$-algebra $\mathscr{Y}$, say corresponding to an observable to be measured in an experiment, the quantum conditional expectation $\mathbb{P}[ X \vert \mathscr{Y}]$ is well-defined provided $X$ commutes with all operators $Y \in \mathscr{Y}$.   The conditional expectation  is the unique operator 
$\mathbb{P}[ X \vert \mathscr{Y}] \in \mathscr{Y}$ such that
\begin{equation}
\mathbb{P} [  X Y ] = \mathbb{P}[ \ \mathbb{P}[ X \vert \mathscr{Y}] Y],
\label{eq:filter-0}
\end{equation}
for any $X \in \mathscr{Y}'  = \{  AY=YA \ \mathrm{for\ all} \ Y  \in \mathscr{Y} \}  $ (the {\em commutant} of $\mathscr{Y}$) and $Y\in \mathscr{Y}$.   It is the orthogonal projection of $\mathscr{Y}'$ onto $\mathscr{Y}$ with respect to the inner product $( A, B ) = \mathbb{P} [ A^\ast B ]$, as in Figure \ref{fig:c-exp-1}. As in the classical case, quantum conditional expectation boils down to {\em least squares estimation}.

\subsection{System-Probe Model for Quantum Filtering}
\label{sec:filter-system-probe}
 
 In the case of the atom coupled to field channels, we cannot access atomic observables directly, and instead we must rely on indirect information available in an output field channel. The interaction between the atom and the field causes information about the atom to be transferred to the field. 
We may monitor an observable of the field, and then  make inferences  about the atom from the data obtained. This is a quantum filtering problem for the atom. The atom-field system is  an instance of the system-probe model of Von Neumann. 
 
In the absence of controls, the atom-field system is defined by the parameters 
\begin{equation}
A = ( \sqrt{\kappa_1}\, \sigma_-, \, \frac{1}{2} \omega \sigma_z   ) \boxplus ( \sqrt{\kappa_2}\, \sigma_-, 0) ,
\label{eq:probe-1}
\end{equation}
which determine the Schrodinger equation
\begin{eqnarray}
d U(t) &=& \{ \sqrt{\kappa_1}\, dB_1^\ast(t) \sigma_-  - \sqrt{\kappa_1}\, \sigma_+ dB_1(t) 
+ \sqrt{\kappa_2}\, dB_2^\ast(t) \sigma_-  - \sqrt{\kappa_2}\, \sigma_+ dB_2(t) 
\nonumber \\ 
&& \hspace{2.0cm} 
-  (\frac{1}{2}  (\kappa_1+\kappa_2)   \sigma_+ \sigma_-   + i \frac{1}{2}\omega \sigma_z    ) dt \} U(t),
\label{eq:probe-2}
\end{eqnarray}
where the two-channel field $B(t)=(B_1(t), B_2(t))^T$ is in the vacuum state $\vert 00 \rangle$.
The output field is defined by $B_{out}(t) = U^\ast(t) B(t) U(t)$, of which we monitor the real quadrature of channel 1, $B_{out,1}(t)+B_{out,1}^\ast(t)$; this  may be achieved by use of an ideal  homodyne detector.

Now the spectral theorem holds in situations  more general than discussed earlier (section \ref{sec:filter-probability}), and in particular may be applied to the algebra of field operators. The field observable $Y(t)=B_{out,1}(t)+B_{out,1}^\ast(t)$ is self-adjoint for each $t$, and for different $t$'s they commute: $[Y(t), Y(t') ]=0$. 
This determines a commutative algebra $\mathscr{Y}_t $ generated by $Y(s)$, for all $0 \leq s \leq t$.
The spectral theorem says that $Y(\cdot)$ is equivalent to a classical stochastic process $\iota(Y)(\cdot)$, the signal generated by the detector, with respect to a classical probability distribution $\mathbf{P}$. Due to their statistical equivalence, we do not usually distinguish between them, and simply write $Y(t)$ for the measurement signal produced by the detector.  

Now let's consider the atom-field system at the initial time $t=0$, before the interaction has taken place. Atomic operators $X \in \mathscr{M}_2$ are represented in the tensor product $\mathscr{M}_2 \otimes \mathscr{F}$ by $X \otimes I$. Field operators $F  \in \mathscr{F}$ are represented by $I \otimes F$. Clearly, these operators commute: $[ X \otimes I , I \otimes F]=0$. If the atom and field are allowed to interact via a unitary $U(t)$, then this commutation relation is preserved: $[ U^\ast(t) (X \otimes I ) U(t) , U^\ast(t) (I \otimes F) U(t) ] = U^\ast(t) [ X \otimes I , I \otimes F] U(t) =0$. This means that in the Heisenberg picture, atomic operators $X(t)$ commute with $Y(t)$.  Indeed, it can be shown that $[X(t), Y(s)]=0$ for all $s \leq t$, i.e. $X(t) \in \mathscr{Y}_t'$.
All of this means that, given a atom-field state $\rho \otimes \vert 00 \rangle \langle 00 \vert$,  the conditional expectation $\hat X(t)=\mathbb{P}[ X(t) \, \vert  \,  \mathscr{Y}_t ]$ is well-defined. 
The conditional expectation $\hat X(t)=\mathbb{P}[ X(t) \, \vert  \,  \mathscr{Y}_t ]$ provides us with an {\em estimate} of $X(t)$ given the measurement signal $Y(s), \, s\leq t$.
The quantum filter computes this conditional expectation, as we will soon  see. 

\subsection{The Quantum Filter}
\label{sec:filter-qf}

In order to present the quantum filter for the atomic system, we need some notation. A normalized {\em conditional state} $\pi_t$ is defined by 
\begin{equation}
\hat X(t)=\pi_t(X) = \mathbb{E}[ X(t) \, \vert \, \mathscr{Y}_t ]
\label{eq:filter-1}
\end{equation}
for any atomic operator $X \in \mathscr{M}_2$. The {\em quantum filter} is a stochastic differential equation for the conditional state:
\begin{eqnarray}
d \pi_t(X) &=&  \pi_t(  -i [X, H] + \mathcal{L}_{L_1}(X) + \mathcal{L}_{L_2}(X)) dt 
\label{eq:filter-2}  \\
&& + (  \pi_t( L_1^\ast X + X L_1) - \pi_t( L^\ast_1 + L_1) \pi_t(X) )
( dY(t) - \pi_t( L_1^\ast + L_1) dt ),
\nonumber 
\end{eqnarray}
with initial condition $\pi_0(X) = \mathrm{tr}[\rho X]$, where $\rho$ is the initial atomic state. Here, we have expressed the filter in terms of the atomic parameters $H=\frac{1}{2}\omega \sigma_z$, $L_1 = \sqrt{\kappa_1}\, \sigma_-$ and $L_2 = \sqrt{\kappa_2}\, \sigma_-$.  
It is quite common in the literature to express the quantum filter in terms of a {\em conditional density operator}  $\hat\rho_t$,  so that $\hat X(t)=\pi_t(X)=\mathrm{tr}[\hat\rho_t X]$. The quantum filter (\ref{eq:filter-2}) takes the form 
\begin{eqnarray}
d \hat\rho_t &=&  ( i[\hat\rho_t, H] + \mathcal{L}^\ast_{L_1}( \hat\rho_t) + \mathcal{L}^\ast_{L_2}( \hat\rho_t))dt
\label{eq:filter-3} \\
&& +
(L_1 \hat\rho_t + \hat\rho_t L_1^\ast - \mathrm{tr} [ (L_1+L_1^\ast) \hat\rho_t] \hat\rho_t )
 ( dY(t)- \mathrm{tr}[ (L_1+L_1^\ast) \hat\rho_t ] dt),
\nonumber 
\end{eqnarray}
with initial condition $\hat\rho_0=\rho$, the initial atomic density operator. Explicitly, if we write
$x (t) = \pi_t(\sigma_x)$, etc, the quantum filter for the atom is
 \begin{eqnarray}
 d x(t) &=&  ( -\omega y(t) - \frac{\kappa_1+\kappa_2}{2} x(t)) dt 
 \nonumber \\
 &&  \hspace{1.0cm} + \sqrt{\kappa_1}\, 
(1+ z(t) -  x^2(t)  ) (dY(t) -  x(t)dt),
 \label{eq:filter-30-x} 
   \\
 d  y(t) &=& (\omega  x(t) - \frac{\kappa_1+\kappa_2}{2}  y(t)) dt 
 \nonumber \\
 && \hspace{2.0cm}  +  
 \sqrt{\kappa_1}\,  x(t)  y(t)  (dY(t) -  x(t)dt) ,
  \label{eq:filter-30-y}
 \\
 d  z(t) &=& (-  (\kappa_1+\kappa_2)  z(t) -(\kappa_1+\kappa_2)  ) dt 
 \nonumber \\
 && \hspace{1.5cm}
 - \sqrt{\kappa_1}\, \ x(t)( 1+ x(t)) (dY(t) -  x(t)dt) .
  \label{eq:filter-30-z}
 \end{eqnarray}

\begin{figure}[h]
\begin{center}
\includegraphics{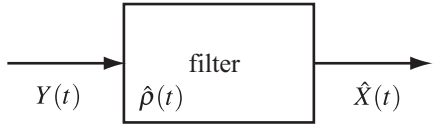}
\caption{The quantum filter produces estimates $\hat X(t)$ from the measurement signal $Y(t)$. The conditional density $\hat{\rho}(t)$ is internal to the filter.
}
\label{fig:filter1}
\end{center}
\end{figure}

The quantum filter has the same form as the classical nonlinear filter due (independently) to Kushner and Stratonovich (see, for example, \cite[Chapter 18]{RE82}).
The filter is driven by the measurement signal $Y(t)$, Figure \ref{fig:filter1}.

As in the classical case, the stochastic process
\begin{equation}
W(t) = Y(t) - \int_0^t \mathrm{tr}[ (L_1+L_1^\ast) \hat\rho_s ] ds 
\label{eq:filter-4}
\end{equation}
is a standard Brownian motion (with respect to the classical probability distribution $\mathbf{P}$ determined by the atom-field state and the measurement observables $\mathscr{Y}_t$), called the {\em innovation process}; it carries the new information available at time $t$. In differential form, we have $dW(t) = dY(t)-\hat x(t) dt$.

The density operator $\rho_t$ for the atom  is  the mean of the conditional density $\hat\rho_t$, since by the fundamental property (\ref{eq:filter-0}) of conditional expectation, we have  for any atomic operator $X$,
\begin{eqnarray}
\mathrm{tr}[ \rho_t X ] &=&
\mathbb{P}[ X(t) ] 
\nonumber \\
&=&
\mathbb{P}[ \pi_t(X) ]
\nonumber \\
&=& \mathbf{P}[ \mathrm{tr}[ \hat{\rho}_t X ] ] .
\label{eq:filter-5}
\end{eqnarray}
Note that in the classical expectation in (\ref{eq:filter-5}), we are averaging the innovation process $W(t)$, a standard Wiener process with respect to $\mathbf{P}$ as mentioned above.

Therefore the {\em master equation} is obtained by averaging the quantum filter:
\begin{equation}
\dot {\rho} = i[\rho, H] +  \mathcal{L}^\ast_{L_1}( \rho) + \mathcal{L}^\ast_{L_2}( \rho),
\label{eq:filter-6}
\end{equation}
with initial condition $\rho_0=\rho$. In terms of coordinates in the Bloch sphere, we have
\begin{eqnarray}
\dot{x}(t) &=& - \omega y(t) - \frac{\kappa_1+\kappa_2}{2} x(t),
\label{eq:filter-60-x} \\
\dot{y}(t) &=& \omega x(t) - \frac{\kappa_1+\kappa_2}{2} y(t),
\label{eq:filter-60-y} \\
\dot{z}(t) &=& -(\kappa_1+\kappa_2) z(t) - (\kappa_1+\kappa_2).
\label{eq:filter-60-z} 
\end{eqnarray}
In   \cite{HC93}, the quantum filter (known as the stochastic master equation) is referred to as an {\em unravelling} of the master equation (\ref{eq:filter-4}).\footnote{The term \lq\lq{unravelling}\rq\rq \  follows from the property $\mathbb{P} [  X  ] = \mathbb{P}[ \ \mathbb{P}[ X \vert \mathscr{Y}] ]$, a consequence of the definition of conditional expectation, (\ref{eq:filter-0}) with $Y=I$). }

\subsection{Derivation of the Quantum Filter}
\label{sec:filter-qf-derivation}

There are several ways of deriving the quantum filter. Here we briefly explain the {\em reference probability method}, which involves rotating  the filtering problem back to the input field. This provides a convenient reference and simplifies calculations.\footnote{This is analogous to the use of a Girsanov transformation to change probability distributions in classical probability.}

To this end, we define the input field quadrature $Z(t) = B_1(t) + B_1^\ast(t)$, and write $\mathscr{Z}_t$ for the {\em commutative} algebra generated by $Z(s)$, $0 \leq s \leq t$. With the input fields in the vacuum state, $Z(t)$ is equivalent to a standard Wiener process.
Now $Y(s) = U^\ast(t) Z(s) U(t)$ for all $s \leq t$, and so we have $\mathscr{Y}_t = U^\ast(t) \mathscr{Z}_t U(t)$, a rotation of the commutative algebras. Next, we rotate the quantum expectation  by defining 
\begin{equation}
\mathbb{Q}_t [ X ] = \mathbb{P}[ U^\ast(t) X U(t) ].
\label{eq:filter-derive-1}
\end{equation}
Now by the definition of conditional expectation, we have 
\begin{equation}
\mathbb{P}[ X(t) \vert \mathscr{Y}_t ] = U^\ast(t) \mathbb{Q}_t [ X \vert \mathscr{Z}_t ] U(t) ,
\label{eq:filter-derive-2}
\end{equation}
and so we need to calculate $\mathbb{Q}_t [ X \vert \mathscr{Z}_t ]$. This may be achieved using a version of Bayes formula, \cite{BH06},  \cite[sec. 3.2]{BHJ07}, and by employing a trick \cite{AH91}, \cite{AH01}. Now $\mathbb{Q}_t$ is defined in terms of $U(t)$, which does not commute with $\mathscr{Z}_t$, and so we replace it by $\tilde U(t) \in \mathscr{Z}_t'$ defined by
\begin{eqnarray}
d \tilde U(t) &=& \{ \sqrt{\kappa_1}\,  \sigma_-  dZ(t) 
+ \sqrt{\kappa_2}\, dB_2^\ast(t) \sigma_-  - \sqrt{\kappa_2}\, \sigma_+ dB_2(t) 
\nonumber \\ 
&& \hspace{2.0cm} 
-  (\frac{1}{2} \kappa^2 \sigma_+ \sigma_-   + i \frac{1}{2}\omega \sigma_z  ) dt \} \tilde U(t),
\label{eq:filter-derive-3}
\end{eqnarray}
with initial condition $\tilde U(0)=I$. Equation (\ref{eq:filter-derive-3}) is almost exactly the same as equation (\ref{eq:probe-2}), except that the coefficient of $dB_1(t)$ has been changed. The justification for this is the fact that, for any atomic state vector $\vert \psi \rangle$, since  $dB_1(t)\vert 00 \rangle=0$,  we have $U(t) \vert \psi \rangle \otimes \vert 00 \rangle = \tilde U(t) \vert \psi \rangle \otimes \vert 00 \rangle$, which ensures $\mathbb{Q}_t [X ] = \mathbb{P}[ \tilde U^\ast (t) X \tilde U(t) ]$. Using the definition of conditional expectations, it can be checked that $\mathbb{P}[ \tilde U^\ast(t) X \tilde U(t) \vert \mathscr{Z}_t ] = \mathbb{P}[ \tilde U^\ast(t) \tilde U(t) \vert \mathscr{Z}_t ]  \mathbb{Q}_t[ X  \vert \mathscr{Z}_t ] $. Therefore
\begin{equation}
\mathbb{Q}_t[ X  \vert \mathscr{Z}_t ] = \frac{ \mathbb{P}[ \tilde U^\ast(t) X \tilde U(t) \vert \mathscr{Z}_t ] }{ \mathbb{P}[ \tilde U^\ast(t) \tilde U(t) \vert \mathscr{Z}_t ] }.
\label{eq:filter-derive-4}
\end{equation}

If we now define an unnormalized conditional expectation
\begin{equation}
 \nu_t(X) = U^\ast(t) \mathbb{P}[ \tilde U^\ast(t) X \tilde U(t) \vert \mathscr{Z}_t ] U(t),
\label{eq:filter-derive-5}
\end{equation}
we see from (\ref{eq:filter-derive-2}) and (\ref{eq:filter-derive-4}) that the normalized conditional expectation is given by 
\begin{equation}
\pi_t(X) = \frac{ \nu_t(X)  }{ \nu_t(I) } .
\label{eq:filter-derive-6}
\end{equation}
Using the quantum Ito rule and conditioning, we find that
\begin{eqnarray}
d \nu_t(X) = \nu_t(  -i [X, H] + \mathcal{L}_{L_1}(X) + \mathcal{L}_{L_2}(X)) dt 
+ \nu_t( L_1^\ast  X + X L_1) dY(t), \hspace{5mm}
\label{eq:filter-derive-7}
\end{eqnarray}
with initial condition $\nu_0(X) = \mathrm{tr}[\rho X]$.
Equation (\ref{eq:filter-derive-7})  is an unnormalized form of the quantum filter (\ref{eq:filter-2}), analogous to
the classical Duncan-Mortensen-Zakai equation (see, for example,  \cite[Chapter 18]{RE82}).

In terms of the unnormalized 
 conditional density $\hat\varrho_t$ (so that $\nu_t(X)=\mathrm{tr}[\hat\varrho_t X]$), we have
\begin{eqnarray}
d \hat\varrho_t &=&  ( i[\hat\varrho_t, H] + \mathcal{L}^\ast_{L_1}( \hat\varrho_t) + \mathcal{L}^\ast_{L_2}( \hat\varrho_t))dt
+ (L_1 \hat\varrho_t + \hat\varrho_t L_1^\ast) dY(t),
\label{eq:filter-derive-8}
\end{eqnarray}
with initial condition $\hat\varrho_0 = \rho$.  Since $\hat\varrho_t$ is not normalized, we augment the representation (\ref{eq:open-3}) for a density matrix   by including the normalization factor $n = \mathrm{tr}[ \hat\varrho]$:
\begin{equation}
\varrho = \frac{1}{2} ( n I + x \sigma_x + y \sigma_y + z \sigma_z ) .
\label{eq:filter-derive-70}
\end{equation}
The unnormalized quantum filter may be expressed in terms of the extended Bloch vector
 $\check r=(n,x,y,z)^T$ as follows:
 \begin{eqnarray}
 d n(t) &=& \sqrt{\kappa_1}\, x(t) dY(t),
  \label{eq:filter-derive-80-n}
  \\ 
 d  x(t) &=&  ( -\omega  y(t) - \frac{\kappa_1+\kappa_2}{2}  x(t)) dt 
+ \sqrt{\kappa_1}\, 
( n(t) + z(t) ) dY(t) ,
 \label{eq:filter-derive-80-x} 
   \\
 d y(t) &=& (\omega  x(t) - \frac{\kappa_1+\kappa_2}{2}  y(t)) dt ,
  \label{eq:filter-derive-80-y}
 \\
 d z(t) &=& (-  (\kappa_1+\kappa_2)  z(t) -(\kappa_1+\kappa_2)n(t)  ) dt 
 - \sqrt{\kappa_1}\, x(t) dY(t)  ,
  \label{eq:filter-derive-80-z}
 \end{eqnarray}
 where here $ x(t) = \nu_t(\sigma_x)$, etc. The normalized quantum filter (\ref{eq:filter-30-x})-(\ref{eq:filter-30-z}) may be obtained from (\ref{eq:filter-derive-80-n})-(\ref{eq:filter-derive-80-z}) by dividing by $n(t)$ and using Ito's rule.

Can we average the unnormalized quantum filter (\ref{eq:filter-derive-8})  and obtain the master equation (\ref{eq:filter-6})? The answer is that we can provided we use the correct expectation. Indeed, define the quantum expectation
\begin{equation}
\mathbb{P}_t^0[ X ] = \mathbb{P}[ U(t) X U^\ast(t) ] .
\label{eq:filter-derive-9}
\end{equation}
Then with respect to $\mathbb{P}_t^0$, the measurement signal $Y(s)$, $0 \leq s \leq t$,  has the same statistics as the input quadrature $Q(t)$ has with respect to $\mathbb{P}$. By the spectral theorem, $Y(t)$ is equivalent to a standard Wiener process with resect to a classical probability distribution $\mathbf{P}^0_t$. Consequently,   for any atomic operator $X$,
\begin{eqnarray}
\mathrm{tr}[ \rho_t X ] &=&
\mathbb{P}[ X(t) ]  = \mathbb{P}[ \mathbb{P} [ \tilde U^\ast(t) X \tilde U(t) ] \vert \mathscr{Z}_t ] ]
\nonumber \\
&=&
\mathbb{P}[ U(t) U^\ast(t) \mathbb{P} [ \tilde U^\ast(t) X \tilde U(t) ] \vert \mathscr{Z}_t ]  U(t) U^\ast(t)]
\nonumber \\
&=&
\mathbb{P}^0_t[ \nu_t(X) ]
\nonumber \\
&=& \mathbf{P}^0_t[ \mathrm{tr}[ \hat{\varrho}_t X ] ] .
\label{eq:filter-derive-10}
\end{eqnarray}
The classical expectation in the last line of (\ref{eq:filter-derive-10}) averages $Y(s)$, $0 \leq s \leq t$,   with respect to $\mathbf{P}^0_t$.

\subsection{Comments}
\label{sec:filter-comments}

In hindsight,  quantum filtering seems to be a natural generalization of classical estimation ideas. It should be understood that the development of the quantum filter was an impressive intellectual achievement, due to a number of authors in the 1980's, notably Belavkin and Carmichael. Certainly, looking at the postulates of quantum mechanics as they are typically presented, it is far from clear how filtering ideas might emerge.
 Quantum filtering builds on the underlying theoretical framework for open quantum systems, which includes quantum operations, master equations and QSDEs, that was developed over several decades, largely in quantum optics. Indeed, filtering ideas have not been developed in other areas of quantum physics, as far as I know, with the   exception of Korotkov's   work in solid state physics (for example, \cite{AK01}).
However, the fundamental statistical notion of conditional expectation is universal when correctly implemented.



\section{Optimal Measurement Feedback Control}
\label{sec:optimal}

In Sections \ref{sec:open-loop-optimal} and \ref{sec:open-loop-impulsive} we saw how optimal control methods could be used to design {\em open loop} control signals. In this section we allow the classical control signals to depend on a measurement signal (in a causal way) and formulate a performance criterion to optimize; the result will be a measurement feedback control system that has been optimally designed. In general, the optimal controller will be a classical dynamical system that processes the measurement signal to produce the control actions. The nature of the controller dynamics will depend on the performance criterion used. In what follows we discuss two performance criteria, known as {\em risk-neutral} and {\em risk-sensitive}, \cite{J04,J05}.
The risk-neutral criterion  leads to an optimal controller whose internal dynamics are given by the quantum filter, \cite{DJ99}, \cite{EB05}. This is quite natural, and generalizes classical results going back to Kalman's  LQG control theory. However, the risk-sensitive criterion gives rise to a different type of filter, first obtained classically by \cite{W81}.

In   this section we consider continuous dynamics of the two-level atom, and omit the impulsive controls. 
The system under control is defined by the parameters 
\begin{equation}
A = ( \sqrt{\kappa_1}\, \sigma_-, \, \frac{1}{2} (\omega \sigma_z   + u \sigma_x ) ) \boxplus ( \sqrt{\kappa_2}\, \sigma_-, 0) ,
\label{eq:direct-1}
\end{equation}
where $u(t)$ is the classical control signal determined by a classical controller $K$ from information in the measurement signal $Y(t)=B_{out,1}(t)+B_{out,1}^\ast(t)$ (recall section \ref{sec:filter}), as in Figure \ref{fig:atom-2} with $\gamma$ empty).

A {\em measurement feedback controller}  $K$ is a causal classical system that processes a measurement signal $Y(s)$, $0 \leq s \leq t$, to produce control actions $u(t)$. We may write $u(t) = K_t( Y(s), \, 0 \leq s \leq t)$.

\subsection{Optimal Risk Neutral  Control}
\label{sec:optimal-rn}

Suppose we wish to maintain the atom in its excited state by measurement feedback. In order do do this in an optimal fashion, we must first encode this objective in a performance criterion $J(K)$ which we  subsequently minimize.  In order to specify $J(K)$, we need to define some cost observables.
We need an observable $C_0 \geq 0$  such that $\langle +1 \vert C_0 \vert +1 \rangle=0$ and $\langle \psi  \vert C_0 \vert \psi \rangle > 0$ for all atomic states $\vert \psi \rangle \neq \vert +1 \rangle$, so that the excited state minimizes the expected value of $C_0$. The choice $C_0 = \vert -1 \rangle \langle -1 \vert=\mathrm{diag}(0,1)$ meets these conditions. We allow the control $u$ to be unbounded, but impose a penalty $c_1 \vert u \vert^2$, for a positive real number $c_1$. Combining, we define the cost observable
\begin{eqnarray}
C_1(u) = C_0 + \frac{c_1}{2} \vert u \vert^2 = \left(  \begin{array}{cc}
\frac{c_1}{2} \vert u \vert^2 & 0
\\
0 & 1+ \frac{c_1}{2}  \vert u \vert^2
\end{array}\right) ,
\label{eq:rn-1}
\end{eqnarray}
which will be integrated along a time interval $[0,T]$. We also define a cost observable for the final time
\begin{equation}
C_2 = c_2 C_0 = \left(  \begin{array}{cc}
0& 0
\\
0 & c_2
\end{array}\right) ,
\label{eq:rn-2}
\end{equation}
where $c_2$ is a positive real number. 

We can now define the performance criterion
\begin{equation}
J(K) = \mathbb{P} \left[  \int_0^T C_1(t) dt + C_2(T) \right],
\label{eq:rn-3}
\end{equation}
where $C_1(t) = U^\ast(t) C_1(u(t)) U(t)$ and $C_2(T) = U^\ast(T) C_2 U(T)$. The risk-neutral optimal control problem is to find a measurement feedback controller $K$ that minimizes $J(K)$, \cite{J05}.

The key step in solving this optimization problem is to re-express the performance criterion $J(K)$ in terms of quantities computable from the measurement signal. The obvious choice is to use the conditional  state $\hat{\rho}_t =\frac{1}{2}(I+x(t)\sigma_x + y(t) \sigma_y + z(t) \sigma_z)$, which is possible because of the fundamental property (\ref{eq:filter-0}) of conditional expectations. Indeed, we have
\begin{eqnarray}
J(K) &=& \mathbb{P} \left[  \int_0^T    \mathbb{P}[ C_1(t) \vert \mathscr{Y}_t ]  dt + \mathbb{P}[ C_2(T)  \vert \mathscr{Y}_T ] \right]
\nonumber \\
&=& \mathbb{P} \left[    \int_0^T \pi_t( C_1(u(t)) ) dt + \pi_T( C_2) \right]
\nonumber \\
&=&
\mathbf{P} \left[   \frac{1}{2}  \int_0^T  (1-z(t) + c_1 \vert u(t) \vert^2) dt + \frac{c_2}{2} (1-z(T)) \right] .
\label{eq:rn-4}
\end{eqnarray}
Now $J(K)$ is expressed in terms of the Bloch vector $r=(x,y,z)^T$, which evolves according to   the controlled quantum filter
  \begin{eqnarray}
 d x(t) &=&  ( -\omega y(t) - \frac{\kappa_1+\kappa_2}{2} x(t)) dt 
 \nonumber \\
 &&  \hspace{1.0cm} + \sqrt{\kappa_1}\, 
(I + z(t) -  x^2(t)  ) (dY(t) -  x(t)dt),
 \label{eq:rn-5-x} 
   \\
 d  y(t) &=& (\omega  x(t) - \frac{\kappa_1+\kappa_2}{2}  y(t)  -u(t)z(t) ) dt 
 \nonumber \\
 && \hspace{2.0cm}  +  
 \sqrt{\kappa_1}\,  x(t)  y(t)  (dY(t) -  x(t)dt) ,
  \label{eq:rn-5-y}
 \\
 d  z(t) &=& (-  (\kappa_1+\kappa_2)  z(t) -(\kappa_1+\kappa_2)  +u(t) y(t) ) dt 
 \nonumber \\
 && \hspace{1.5cm}
 - \sqrt{\kappa_1}\,  x(t)( 1+ x(t)) (dY(t) -  x(t)dt) .
  \label{eq:rn-5-z}
 \end{eqnarray}
 
 Equations (\ref{eq:rn-5-x})-(\ref{eq:rn-5-z}) are driven by the measurement data $Y(t)$, and so the conditional state $r(t)=(x(t), y(t), z(t))^T$ is available to the controller. Also, the innovations process $dW(t)=dY(t)-x(t)dt$ is a standard Wiener process, independent of the controller, and so we may regard equations (\ref{eq:rn-5-x})-(\ref{eq:rn-5-z}) as being driven by $W(t)$. Therefore we may re-write  (\ref{eq:rn-5-x})-(\ref{eq:rn-5-z})  and the performance criterion $J(K)$ in the compact forms
 \begin{equation}
 d r(t) = f(r(t), u(t) ) dt + g(r(t) ) dW(t) ,
 \label{eq:rn-6}
 \end{equation}
 and
  \begin{equation}
J(K) = \mathbf{P} \left[ \int_0^T L(r(t), u(t)) dt + M(r(T)) \right] ,
 \label{eq:rn-7}
 \end{equation}
 where now the controller $K$ determines $u(t)$  causally from knowledge of $r(s)$, $0 \leq s \leq t$.
 Hence
we 
have converted the original quantum measurement feedback optimal control problem into an equivalent classical stochastic control problem with full state information. This equivalent problem may be solved using standard methods of classical stochastic control theory,  \cite{FR75}, \cite{FS06}.  

Let's apply dynamic programming. The value function is defined by
\begin{equation}
V(r,t) = \inf_{u(\cdot)}  
\mathbf{P} \left[
\int_t^T L(r(s), u(s)) ds + M(r(T)) \ : \ r(t)=r
\right],
\label{eq:rn-8}
 \end{equation}
 where (\ref{eq:rn-6})  is initialized at the Bloch vector  $r(t)=r$, and the infimum is over open loop controls $u(\cdot)$.
  The dynamic programming equation is
 \begin{eqnarray}
 \frac{\partial}{\partial t} S(r,t) + \min_{u} \{ \mathcal{L}^u S(r,t) + \frac{1}{2} ( 1-z+c_1 \vert u \vert^2 ) \} &=& 0,
 \label{eq:rn-9-a} \\
 S(r,T) &=& M(r) ,
\label{eq:rn-9-b}
\end{eqnarray}
where $\mathcal{L}^u$ is the generator of the SDE (\ref{eq:rn-6}): for a smooth function $\varphi(r)$, 
\begin{eqnarray}
\mathcal{L}^u \varphi(r) = \frac{1}{2} \mathrm{tr} [g(r) g(r)^T  D^2 \varphi(r)  ] + D\varphi(r)[ f(r,u) ] .
\label{eq:rn-10}
\end{eqnarray}
Here, $D^2V$ denotes the Hessian matrix of second-order partial derivatives.
The minimum in (\ref{eq:rn-9-a}) is attained at the value
 \begin{equation}
\mathbf{u}^\star(r,t)  = \mathbf{u}^\star(x,y,z,t) =  \frac{1}{c_1}( S_y (x,y,z,t) z -S_z(x,y,z,t) y ).
\label{eq:rn-12}
\end{equation}
By the verification theorem, if we have a smooth solution $S(r,t)$ to the DPE (\ref{eq:rn-9-a})-(\ref{eq:rn-9-b}), and a control $u^\star(t)$ such that
\begin{equation}
u^\star(t)  = \mathbf{u}^\star (x(t), y(t), z(t), t) ,
\label{eq:rn-11}
\end{equation}
then $u^\star(t)$ is optimal and $S(r,t)=V(r,t)$. The expression (\ref{eq:rn-11})  defines an optimal state feedback controller for the equivalent classical problem.

We can now define an optimal measurement feedback controller $K^\star$ for the quantum optimal control problem using the function $\mathbf{u}^\star(r,t)$ defined by (\ref{eq:rn-11}) and the controlled quantum filter   (\ref{eq:rn-5-x})-(\ref{eq:rn-5-z}):
\begin{equation}
 K^\ast_t( Y(\cdot) ) = \mathbf{u}^\star(r(t), t). 
\label{eq:rn-13}
\end{equation}
The master equation for the optimal measurement feedback system may be obtained by substituting (\ref{eq:rn-12}) into (\ref{eq:rn-5-x})-(\ref{eq:rn-5-z}) and taking expectations:
  \begin{eqnarray}
 \dot x(t) &=&    -\omega y(t) - \frac{\kappa_1+\kappa_2}{2} x(t) ,
 \label{eq:rn-14-x} 
   \\
 d  y(t) &=& \omega  x(t) - \frac{\kappa_1+\kappa_2}{2}  y(t)  -  \frac{1}{c_1} S_y z^2(t) + \frac{1}{c_1} S_z z(t) y(t), 
  \label{eq:rn-14-y}
 \\
 d  z(t) &=& -  (\kappa_1+\kappa_2)  z(t) -(\kappa_1+\kappa_2)  +\frac{1}{c_1} S_y z(t) y(t) -\frac{1}{c_1} S_z y^2(t). 
    \label{eq:rn-14-z}
 \end{eqnarray}
 
Examination of the mean closed loop dynamics (\ref{eq:rn-14-x})-(\ref{eq:rn-14-z}) reveals that the equilibria depend on the partial derivatives of the value function $S(x,y,z)$. A complete analysis of the performance of this system would benefit from a detailed numerical study.

Recently, optimal  risk-neutral measurement feedback methods (LQG) were applied to the problem of frequency locking of an optical cavity 
and the control system was experimentally demonstrated, \cite{HHHPJ09}.




\subsection{Optimal Risk Sensitive  Control}
\label{sec:optimal-rs}

In order to define a risk-sensitive performance criterion, we introduce $R(t)$ as the solution to the equation
\begin{equation}
\frac{dR(t)}{dt} = \frac{\mu}{2} C_1(t) R(t),
\label{eq:rs-1}
\end{equation}
with initial condition $R(0)=I$. Here, $C_1(t) = U^\ast(t) C_1(u(t)) U(t)$ as in the previous section, and $\mu > 0$ is a positive real number called a risk parameter. The solution to (\ref{eq:rs-1}) can be expressed as a time-ordered exponential
\begin{equation}
R(t) = \stackrel{\leftarrow}{\exp} \left(\frac{\mu}{2} \int_0^t C_1(s) ds \right) .
\label{eq:rs-2}
\end{equation}
We then define the {\em risk-sensitive} cost function to be the quantum
expectation
\begin{equation}
J^\mu(K) = \mathbb{P}[R^\ast(T) e^{\mu C_2(T)} R(T) ],
\label{eq:rs-3}
\end{equation}
where $C_2(T)= U^\ast(T) C_2 U(T)$. The risk-sensitive optimal control problem is to find a measurement feedback controller $K$ that minimizes $J^\mu(K)$, \cite{J05}.

It can be seen that the risk-sensitive performance criterion $J^\mu(K)$, equation (\ref{eq:rs-3}), has a multiplicative form, in contrast to the additive form used in the risk-neutral criterion $J(K)$, equation (\ref{eq:rn-3}). This multiplicative form precludes us from expressing $J^\mu(K)$ in a useful way in terms of the condition density $\hat{\rho}_t$.  In 1981, Whittle showed how to solve a classical risk-sensitive problem by using a {\em modified} conditional state, with a corresponding modified filter, \cite{W81}, \cite{BV85}, \cite{JBE94}. We now explain how this works for the quantum risk-sensitive criterion $J^\mu(K)$, \cite{J05}, \cite{WDDJ06}.

We begin by noting that $e^{\mu C_2(T)} = U^\ast(T) e^{\mu C_2} U(T)$, and so, in view of the form of the criterion  (\ref{eq:rs-3}), it is natural to define $U^\mu(t) = U(t) R(t)$, which satisfies the QSDE
\begin{eqnarray}
d U^\mu (t) &=& \{ \sqrt{\kappa_1}\, dB_1^\ast(t) \sigma_-  - \sqrt{\kappa_1}\, \sigma_+ dB_1(t) 
+ \sqrt{\kappa_2}\, dB_2^\ast(t) \sigma_-  - \sqrt{\kappa_2}\, \sigma_+ dB_2(t) 
\nonumber \\ 
&& \hspace{0.1cm} 
-  (\frac{1}{2} (\kappa_1 + \kappa_2 ) \sigma_+ \sigma_-   + i \frac{1}{2}\omega \sigma_z  - C_1(u(t)) )  dt \} U^\mu (t),
\label{eq:rs-4}
\end{eqnarray}
with initial condition $U^\mu (0)=I$. Equation (\ref{eq:rs-4}) is a modification of the Schrodinger equation (\ref{eq:probe-2}),  via the inclusion of  the cost observable $C_1(u(t))$. The risk-sensitive performance criterion  $J^\mu(K)$ is therefore 
\begin{equation}
J^\mu(K) = \mathbb{P}[ U^{\mu\, \ast}(T) e^{\mu C_2} U^\mu(T) ].
\label{eq:rs-5}
\end{equation}

Now $U^\mu(t)$ does not commute with $\mathscr{Z}_t$, and so we follow the approach taken to derive the quantum filter in section \ref{sec:filter-qf-derivation}. Define $\tilde U^\mu(t)$ to be the solution of
\begin{eqnarray}
d \tilde U^\mu(t) &=& \{ \sqrt{\kappa_1}\,  \sigma_-  dZ(t) 
+ \sqrt{\kappa_2}\, dB_2^\ast(t) \sigma_-  - \sqrt{\kappa_2}\, \sigma_+ dB_2(t) 
\nonumber \\ 
&& \hspace{0.1cm} 
-  (\frac{1}{2}( \kappa_1+\kappa_2) \sigma_+ \sigma_-   + i \frac{1}{2}\omega \sigma_z    - C_1(u(t)) ) dt \} \tilde U^\mu (t),
\label{eq:rs-6}
\end{eqnarray}
with initial condition $\tilde U^\mu(0)=I$. Then $\tilde V(t)$  commutes with $\mathscr{Z}_t$, and
\begin{equation}
J^\mu(K) = \mathbb{P}[ \tilde U^{\mu\, \ast}(T) e^{\mu C_2} \tilde U^\mu(T) ].
\label{eq:rs-7}
\end{equation}
Then by a calculation similar to (\ref{eq:filter-derive-10}), with $\tilde U^\mu$ replacing $\tilde U$, we find that
\begin{eqnarray}
J^\mu(K) &=&  \mathbb{P}^0_t[  \nu^\mu_t (e^{\mu C_2} )]
\nonumber \\
&=&
\mathbf{P}^0_t [ \mathrm{tr}[ \hat{\varrho}^\mu_t ( e^{\mu C_2} ) ],
\label{eq:rs-8}
\end{eqnarray}
where we introduce the risk-sensitive conditional state
\begin{equation}
 \nu^\mu_t(X) = U^\ast(t) \mathbb{P}[ \tilde U^{\mu \, \ast}(t) X \tilde U^\mu(t) \vert \mathscr{Z}_t ] U(t),
\label{eq:rs-9}
\end{equation}
and corresponding risk-sensitive conditional density: $\nu^\mu_t(X)= \mathrm{tr}[ \hat{\varrho}^\mu_t X]$.

The risk-sensitive quantum filter is
\begin{eqnarray}
d \hat\varrho^\mu_t &=&  ( i[\hat\varrho^\mu_t, H(u(t)] 
+\mathcal{C}^\ast_{C_1(u(t))}(\hat\varrho^\mu_t) 
+ \mathcal{L}^\ast_{L_1}( \hat\varrho^\mu_t) + \mathcal{L}^\ast_{L_2}( \hat\varrho^\mu_t))dt
\nonumber \\
& & \hspace{1.0cm} + (L_1 \hat\varrho^\mu_t + \hat\varrho^\mu_t L_1^\ast) dY(t),
\label{eq:rs-10}
\end{eqnarray}
with initial condition $\hat\varrho^\mu_0 = \rho$, where the running cost superoperator is defined by
\begin{equation}
\mathcal{C}^\ast_{C}(\rho) =  \frac{\mu}{2}( C \rho + \rho C ) .
\label{eq:rs-11}
\end{equation}
In equation (\ref{eq:rs-10}), the measurement signal $Y(s)$, $0\leq s \leq t$,  is a standard Wiener process with respect to the classical probability distribution $\mathbf{P}^0_t$.

The risk-sensitive cost $J^\mu(K)$ may be expressed in terms of the
extended Bloch vector
 $\check r=(n,x,y,z)^T$ by substituting the expression $\hat\rho^\mu= \frac{1}{2} ( n I + x \sigma_x + y \sigma_y + z \sigma_z )$ into (\ref{eq:rs-8}):
\begin{equation}
 J^\mu(K) = \mathbf{P}^0 \left[ 
 \frac{1}{2}( N(T)-z(T)) e^{\mu c_2} 
 \right],
\label{eq:rs-13}
\end{equation}
where
 \begin{eqnarray}
 d n(t) &=& 
 \frac{\mu}{2} (n(t)-z(t) + c_1 \vert u(t) \vert^2 n(t)) dt +
 \sqrt{\kappa_1}\, x(t) dY(t),
  \label{eq:rs-12-n}
  \\ 
 d  x(t) &=&  ( -\omega  y(t) - \frac{\kappa_1+\kappa_2}{2}  x(t)
 +\frac{\mu}{2} (x(t) + c_1 \vert u(t) \vert^2 x(t))
 ) dt 
  \nonumber \\
 && \hspace{1.0cm}
+ \sqrt{\kappa_1}\, 
( n(t) + z(t) ) dY(t) ,
 \label{eq:rs-12-x} 
   \\
 d y(t) &=& (\omega  x(t) - \frac{\kappa_1+\kappa_2}{2}  y(t)  -u(t) z(t)
 + \frac{\mu}{2} (y(t) +  c_1 \vert u(t) \vert^2y(t) )
 ) dt ,
  \label{eq:rs-12-y}
 \\
 d z(t) &=& (-  (\kappa_1+\kappa_2)  z(t) -(\kappa_1+\kappa_2)n(t) + u(t) y(t) 
 -\frac{\mu}{2}( n(t)-z(t) +c_1 \vert u(t) \vert^2 z(t))
 ) dt 
  \nonumber \\
 && \hspace{1.0cm}
 - \sqrt{\kappa_1}\, x(t) dY(t)  .
  \label{eq:rs-12-z}
 \end{eqnarray}
 These equations for $\check r(t)$ are of the form
 \begin{equation}
 \dot{ \check{r}} (t) = f^\mu( \check r(t), u(t)) dt + g^\mu(\check r(t) ) dY(t) .
 \label{eq:rs-15}
 \end{equation}

 The value function for the risk-sensitive problem is defined by
 \begin{equation}
 V^\mu(\check r, t) = \inf_{u(\cdot)} \mathbf{P}^0 \left[
  \frac{1}{2}( N(T)-z(T)) e^{\mu c_2}  \ : \ \check{r}(t) = \check r
 \right],
 \label{eq:rs-14}
 \end{equation}
 where $\check r(\cdot)$ evolves on the time interval $[t,T]$ according to (\ref{eq:rs-15}) with initial condition $\check r(t)= \check r$.
 The corresponding dynamic programming equation is
  \begin{eqnarray}
 \frac{\partial}{\partial t} S^\mu(\check r,t) + \min_{u}  \mathcal{L}^{\mu,u} S^\mu(\check r,t)   &=& 0,
 \label{eq:rs-15-a} \\
 S^\mu(\check r,T) &=&  \frac{1}{2}( n-z)e^{\mu c_2}
\label{eq:rs-15-b}
\end{eqnarray}
where $\mathcal{L}^{\mu,u}$ is the generator of the SDE (\ref{eq:rs-15}): for a smooth function $\varphi(r)$, 
\begin{eqnarray}
\mathcal{L}^{\mu,u} \varphi(r) = \frac{1}{2} \mathrm{tr} [g^\mu(\check r) g^\mu(\check r)^T  D^2 \varphi(\check r)  ] + D\varphi(\check r)[ f^\mu(\check r,u) ] .
\label{eq:rs-16}
\end{eqnarray}
 Now the minimum in (\ref{eq:rs-15-a}) is attained at the value $\mathbf{u}^{\mu, \star}(\check r,t)  = \mathbf{u}^{\mu, \star}(n,x,y,z,t) $ given by
 \begin{eqnarray}
  \mathbf{u}^{\mu, \star}(n,x,y,z,t) =  \frac{1}{\mu c_1 S^\mu(n,x,y,z) }( S^\mu_y (n,x,y,z,t) z -S^\mu_z(n,x,y,z,t) y ).
  \hspace{0.8cm}
\label{eq:rs-17}
\end{eqnarray}
In deriving (\ref{eq:rs-17}), we have used the relation $DS^\mu(\check r,t) \cdot \check r = S^\mu(\check r, t)$ which follows from the multiplicative homogeneity property $V^\mu( \alpha \check r, t)= \alpha V^\mu(\check r, t)$ enjoyed by the value function.

The verification theorem for the risk-sensitive problem asserts that if we have a smooth solution $S^\mu(\check r,t)$ to the DPE (\ref{eq:rs-15-a})-(\ref{eq:rs-15-b}), and a control $u^{\mu, \star }(t)$ such that
\begin{equation}
u^{\mu, \star} (t)  = \mathbf{u}^{\mu, \star} (n(t), x(t), y(t), z(t), t) ,
\label{eq:rs-18}
\end{equation}
then $u^{\mu, \star} (t)$ is optimal and $S^\mu(\check r,t)=V^\mu(\check r,t)$.  

We can now define an optimal measurement feedback controller $K^{\mu,\star}$ for the quantum risk-sensitive optimal control problem using the function $\mathbf{u}^{\mu, \star}(\check r,t)$ defined by (\ref{eq:rs-18}) and the risk-sensitive quantum filter   (\ref{eq:rs-12-x})-(\ref{eq:rs-12-z}):
\begin{equation}
 K^{\mu, \ast}_t( Y(\cdot) ) = \mathbf{u}^{\mu, \star}(\check r(t), t). 
\label{eq:rs-19}
\end{equation}




\subsection{Optimal Measurement Feedback  Impulsive Control}
\label{sec:optimal-rn-impulse}

It should be apparent to the reader that one may formulate and solve optimal measurement feedback problems for systems with impulsive controls. Due to space limitations, we do not pursue this further in this article.

\section{Coherent Feedback Control}
\label{sec:coherent}

As discussed in Section \ref{sec:types}, coherent feedback systems preserve quantum information through the use of a controller which is itself a quantum system, and one or more means of transferring quantum information between the plant and the controller. While the idea of coherent feedback is natural, to date there is little known about how to {\em design} coherent feedback systems in a systematic manner. In this section we discuss two simple examples.

\subsection{Coherent Feedback Control using Direct Couplings}
\label{sec:coherent-direct}

In this section we take a look at how quantum information may be transfered from a quantum controller to a quantum plant using an impulse implementing a rapid coherent interaction between the two systems. This example is based on the coherent spin control example discussed in \cite[Sec. III.E]{SL00}. 

The plant $P$ and controller $C$ are independent  two level systems, with Pauli matrices $\sigma_\alpha^{(P)}$,  $\sigma_\alpha^{(C)}$ ($\alpha=x,y,z$). By suitable choice of reference or otherwise, we assume that these systems have trivial self-energies, and are decoupled from external fields. 
The parameters $P=(0,0)$ and $C=(0,0)$ describe the trivial dynamics $\dot \sigma_\alpha^{(P)}(t)=0$,  $\dot \sigma_\alpha^{(C)}(t)=0$ of these systems in the absence of interaction. 
We assume that interactions between the plant and controller may be described by the action of  bipartite unitaries
\begin{equation}
\mathbf{V} = \{  CNOT_{PC}, \ \  CNOT_{CP} \} .
\label{eq:cfb-d-1}
\end{equation}
Here, $CNOT_{AB}$ is the CNOT gate with $A$ as the control bit:
\begin{eqnarray*}
CNOT_{AB}\vert 00 \rangle &=& \vert 00 \rangle
\\
CNOT_{AB}\vert 01 \rangle &=& \vert 01 \rangle
\\
CNOT_{AB}\vert 10 \rangle &=& \vert 11 \rangle
\\
CNOT_{AB}\vert 11 \rangle &=& \vert 10 \rangle .
\end{eqnarray*}
The interactions are applied via an impulsive control sequence $\gamma$, with unitaries  $V_k \in \mathbf{V}$.
In the Heisenberg picture, the hybrid equations of motion are simply
\begin{eqnarray}
\dot \sigma_\alpha^{(P)}(t) &= & 0, \ \ \tau_k < t \leq \tau_{k+1},
\label{eq:cfb-d-2-P}
\\
\dot \sigma_\alpha^{(C)}(t) &= & 0, \ \ \tau_k < t \leq \tau_{k+1},
\label{eq:cfb-d-2-C}
\\
\sigma_\alpha^{(P)}(\tau_k^+) &=& V_k^\ast (\tau_k)  \sigma_\alpha^{(P)}(\tau_k)  V_k (\tau_k)  ,
\label{eq:cfb-d-2-P-tau}
\\
\sigma_\alpha^{(C)}(\tau_k^+) &=& V_k^\ast (\tau_k)  \sigma_\alpha^{(C)}(\tau_k)  V_k  (\tau_k)  .
\label{eq:cfb-d-2-C-tau}
\end{eqnarray}

The control objective considered in \cite[sec. III.E]{SL00} was to put the plant in the state $\vert \downarrow_P \rangle$. If the plant is in an arbitrary pure initial state, then this objective may be achieved by first initializing the controller in the state  $\vert \downarrow_C \rangle$, and then applying the impulsive control $\gamma = ( (0, CNOT_{PC}), (1, CNOT_{CP}))$, as the reader may readily verify. 

 If the plant and controller are subject to non-trivial dynamics between the application of impulses, such as decoherence effects, then a more general  hybrid dynamical model may be developed along the lines discussed in Section \ref{sec:types}.

\subsection{Coherent Feedback Control using Quantum Signals}
\label{sec:coherent-signals}

In Section \ref{sec:networks} we described a class of quantum feedback networks involving the interconnection of systems or subsystems via freely traveling quantum fields (quantum signals). The examples discussed illustrate the point that this type of feedback may be used to change the dynamical behavior of the plant. An important challenge for control theory is to develop ways of {\em designing} signal-based coherent feedback systems in order to meet performance specifications., \cite{YK03a}, \cite{YK03b}, \cite{JNP08}, \cite{HM08}, \cite{NJP09}, \cite{GJ09}, \cite{KNPM09}, \cite{NJD09}, \cite{JG10}, \cite{HM12}, \cite{CTSAM13}.

\begin{figure}[h]
\begin{center}
\includegraphics{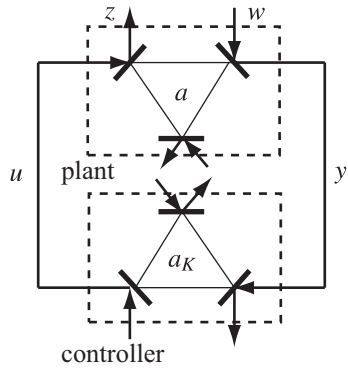}
\caption{Coherent feedback control example, showing plant $a$ and controller  $a_K$ cavity modes,
together with performance quantity $z$ and the \lq\lq{disturbance}\rq\rq \ input $w$.  The coherent signals $u$ and $y$ are used to transfer quantum information between the plant and the controller. The feedback system was designed to minimize the intensity of the  light at the output $z$ when an optical signal is applied at the input  $w$.}
\label{fig:hinfty1}
\end{center}
\end{figure}

While a detailed discussion of signal-based coherent feedback control design is beyond the scope of this article,  we briefly describe an example from \cite{JNP08}, \cite{HM08}. In this example, the plant is a cavity with three mirrors defining three field channels. The problem was to design a coherent feedback system to minimize the influence of one input channel $w$ on an output channel $z$, Figure \ref{fig:hinfty1}.  That is, if light is shone onto the mirror corresponding to the input  channel $w$, we would like the output channel $z$ to be dark. This is a simple example of robust control, where $z$ may be regarded as a performance quantity (to be minimized in magnitude), while $w$ plays the role of an external disturbance.

In \cite{JNP08}, it was shown how such problems could be solved systematically by extending methods from classical robust control theory, and importantly, taking into account the physical realization of the coherent controller as a quantum system. Indeed, the controller designed turned out to be another cavity, with mirror transmissivity parameters determined using mathematical methods.  This approached was validated by experiment \cite{HM08}.


\bibliographystyle{plain}


\begin{thebibliography}{10}

\bibitem{AM79}
B.D.O. Anderson and J.B. Moore.
\newblock {\em Optimal Filtering}.
\newblock Prentice-Hall, Englewood Cliffs, NJ, 1979.

\bibitem{VPB79}
V.~P. Belavkin.
\newblock Optimal measurement and control in quantum dynamical systems.
\newblock preprint 411, {Institute of Physics, Nicolaus Copernicus University,
  Torun}, 1979.

\bibitem{VPB92a}
V.P. Belavkin.
\newblock Quantum stochastic calculus and quantum nonlinear filtering.
\newblock {\em J. Multivariate Analysis}, 42:171--201, 1992.

\bibitem{BV85}
A.~Bensoussan and J.H. van Schuppen.
\newblock Optimal control of partially observable stochastic systems with an
  exponential-of-integral performance index.
\newblock {\em SIAM Journal on Control and Optimization}, 23:599--613, 1985.

\bibitem{BH06}
L.~Bouten and R.~{van~Handel}.
\newblock On the separation principle of quantum control.
\newblock In M.~Guta, editor, {\em Proceedings of the 2006 QPIC Symposium}.
  World Scientific, math-ph/0511021 2006.

\bibitem{BHJ07}
L.~Bouten, R.~{van~Handel}, and M.R. James.
\newblock An introduction to quantum filtering.
\newblock {\em SIAM J. Control and Optimization}, 46(6):2199--2241, 2007.

\bibitem{HC93}
H.~Carmichael.
\newblock {\em An Open Systems Approach to Quantum Optics}.
\newblock Springer, Berlin, 1993.

\bibitem{HJC93}
H.J. Carmichael.
\newblock Quantum trajectory theory for cascaded open systems.
\newblock {\em Phys. Rev. Lett.}, 70(15):2273--2276, 1993.

\bibitem{DD07}
D.~D'Alessandro.
\newblock {\em Introduction to Quantum Control and Dynamics}.
\newblock Chapman and Hall/CRC, 2007.

\bibitem{DJ99}
A.C. Doherty and K.~Jacobs.
\newblock Feedback-control of quantum systems using continuous
  state-estimation.
\newblock {\em Phys. Rev. A}, 60:2700, 1999.

\bibitem{EB05}
S.~C. Edwards and V.~P. Belavkin.
\newblock Optimal quantum feedback control via quantum dynamic programming.
\newblock quant-ph/0506018, University of Nottingham, 2005.

\bibitem{RE82}
R.J. Elliott.
\newblock {\em Stochastic Calculus and Applications}.
\newblock Springer Verlag, New York, 1982.

\bibitem{EJ89}
L.C. Evans and M.R. James.
\newblock The {H}amilton-{J}acobi-{B}ellman equation for time-optimal control.
\newblock {\em SIAM J. Control and Optim.}, 27(6):1477--1489, 1989.

\bibitem{FR75}
W.H. Fleming and R.W. Rishel.
\newblock {\em Deterministic and Stochastic Optimal Control}.
\newblock Springer Verlag, New York, 1975.

\bibitem{FS06}
W.H. Fleming and H.M. Soner.
\newblock {\em Controlled Markov Processes and Viscosity Solutions}.
\newblock Springer Verlag, New York, second edition, 2006.

\bibitem{CWG93}
C.W. Gardiner.
\newblock Driving a quantum system with the output field from another driven
  quantum system.
\newblock {\em Phys. Rev. Lett.}, 70(15):2269--2272, 1993.

\bibitem{GC85}
C.W. Gardiner and M.J. Collett.
\newblock Input and output in damped quantum systems: Quantum stochastic
  differential equations and the master equation.
\newblock {\em Phys. Rev. A}, 31(6):3761--3774, 1985.

\bibitem{GZ00}
C.W. Gardiner and P.~Zoller.
\newblock {\em Quantum Noise}.
\newblock Springer, Berlin, 2000.

\bibitem{GJ09}
J.~Gough and M.R. James.
\newblock The series product and its application to quantum feedforward and
  feedback networks.
\newblock {\em IEEE Trans. Automatic Control}, 54(11):2530--2544, 2009.

\bibitem{GJN09a}
J.~Gough, M.R. James, and H.~Nurdin.
\newblock Linear quantum feedback networks with squeezing components.
\newblock In {\em Proc. 48th IEEE Conference on Decision and Control},
  Shanghai, China, December 2009.

\bibitem{HHHPJ09}
S.Z.~Sayed Hassen, M.~Heurs, E.H. Huntington, I.R. Petersen, and M.R. James.
\newblock Frequency locking of an optical cavity using linear quadratic
  gaussian integral control.
\newblock {\em J. Phys. B: At. Mol. Opt. Phys.}, 42:175501, 2009.

\bibitem{AH91}
A.~Holevo.
\newblock Quantum stochastic calculus.
\newblock {\em J. Soviet Math.}, 56:2609--2624, 1991.

\bibitem{AH01}
A.S. Holevo.
\newblock {\em Statistical Structure of Quantum Theory}.
\newblock Springer, Berlin, 2001.

\bibitem{HTC83}
G.M. Huang, T.J. Tarn, and J.W. Clark.
\newblock On the controllability of quantum-mechanical systems.
\newblock {\em J. Math. Phys.}, 24(11):2608--2618, 1983.

\bibitem{HP84}
R.L. Hudson and K.R. Parthasarathy.
\newblock Quantum {I}to's formula and stochastic evolutions.
\newblock {\em Commun. Math. Phys.}, 93:301--323, 1984.

\bibitem{J04}
M.R. James.
\newblock Risk-sensitive optimal control of quantum systems.
\newblock {\em Phys. Rev. A}, 69:032108, 2004.

\bibitem{J05}
M.R. James.
\newblock A quantum {L}angevin formulation of risk-sensitive optimal control.
\newblock {\em J. Optics B: Semiclassical and Quantum, {S}pecial {I}ssue on
  {Q}uantum {C}ontrol}, 7(10):S198--S207, 2005.

\bibitem{JBE94}
M.R. James, J.S. Baras, and R.J. Elliott.
\newblock Risk-sensitive control and dynamic games for partially observed
  discrete-time nonlinear systems.
\newblock {\em IEEE Transactions on Automatic Control}, 39:780--792, 1994.

\bibitem{JG10}
M.R. James and J.~Gough.
\newblock Quantum dissipative systems and feedback control design by
  interconnection.
\newblock {\em IEEE Trans Auto. Control}, 55(8):1806--1821, August 2010.

\bibitem{JNP08}
M.R. James, H.~Nurdin, and I.R. Petersen.
\newblock ${H}^\infty$ control of linear quantum systems.
\newblock {\em IEEE Trans Auto. Control}, 53(8):1787--1803, 2008.

\bibitem{REK60a}
R.E. Kalman.
\newblock Contributions to the theory of optimal control.
\newblock {\em Boletin de la Sociedad Matematica Mexicana}, 5:102--119, 1960.

\bibitem{KNPM09}
J.~Kerckhoff, H.I. Nurdin, D.S. Pavlichin, and H.~Mabuchi.
\newblock Coherent-feedback formulation of continuous quantum error correction
  protocol.
\newblock arxiv:0907.0236, 2009.

\bibitem{KBG01}
N.~Khaneja, R.~Brockett, and S.J. Glaser.
\newblock Time optimal control in spin systems.
\newblock {\em Phys. Rev. A}, (63):032308, 2001.

\bibitem{AK01}
A.N. Korotkov.
\newblock Selective quantum evolution of a qubit state due to continuous
  measurement.
\newblock {\em Phys. Rev. B}, 63:115403, 2001.

\bibitem{SL00}
S.~Lloyd.
\newblock Coherent quantum feedback.
\newblock {\em Phys. Rev. A}, 62:022108, 2000.

\bibitem{HM08}
H.~Mabuchi.
\newblock Coherent-feedback quantum control with a dynamic compensator.
\newblock {\em Phys. Rev. A}, 78(3):032323, 2008.

\bibitem{NJD09}
H.~Nurdin, M.R. James, and A.C. Doherty.
\newblock Network synthesis of linear dynamical quantum stochastic systems.
\newblock {\em SIAM J. Control and Optim.}, 48(4):2686--2718, 2009.

\bibitem{NJP09}
H.~Nurdin, M.R. James, and I.R. Petersen.
\newblock Coherent quantum {LQG} control.
\newblock {\em Automatica}, 45:1837--1846, 2009.

\bibitem{KRP92}
K.R. Parthasarathy.
\newblock {\em An Introduction to Quantum Stochastic Calculus}.
\newblock Birkhauser, Berlin, 1992.

\bibitem{RVMK00}
H.~Rabitz, R.~de~Vivie-Riedle, M.~Motzkus, and K.~Kompa.
\newblock Whither the future of controlling quantum phenomena?
\newblock {\em Science}, 288:824--828, 2000.

\bibitem{SJ08}
S.~Sridharan and M.R. James.
\newblock Minimum time control of spin systems via dynamic programming.
\newblock In {\em Proc. IEEE CDC}, 2008.

\bibitem{W81}
P.~Whittle.
\newblock Risk-sensitive linear/ quadratic/ {G}aussian control.
\newblock {\em Advances in Applied Probability}, 13:764--777, 1981.

\bibitem{JW97}
J.C. Willems.
\newblock On interconnections, control, and feedback.
\newblock {\em IEEE Trans. Automatic Control}, 42(3):326--339, March 1997.

\bibitem{WDDJ06}
S.D. Wilson, C.~D'Helon, A.C. Doherty, and M.R. James.
\newblock Quantum risk-sensitive control.
\newblock In {\em Proc. 45th IEEE Conference on Decision and Control}, pages
  3132--3137, December 2006.

\bibitem{WM94b}
H.~M. Wiseman and G.~J. Milburn.
\newblock All-optical versus electro-optical quantum-limited feedback.
\newblock {\em Phys. Rev. A}, 49(5):4110--4125, 1994.

\bibitem{WM10}
H.M. Wiseman and G.J. Milburn.
\newblock {\em Quantum Measurement and Control}.
\newblock Cambridge University Press, Cambridge, UK, 2010.

\bibitem{YK03a}
M.~Yanagisawa and H.~Kimura.
\newblock Transfer function approach to quantum control-part {I}: Dynamics of
  quantum feedback systems.
\newblock {\em IEEE Trans. Automatic Control}, (48):2107--2120, 2003.

\bibitem{YK03b}
M.~Yanagisawa and H.~Kimura.
\newblock Transfer function approach to quantum control-part {II}: Control
  concepts and applications.
\newblock {\em IEEE Trans. Automatic Control}, (48):2121--2132, 2003.

\bibitem{YD84}
B.~Yurke and J.S. Denker.
\newblock Quantum network theory.
\newblock {\em Phys. Rev. A}, 29(3):1419--1437, 1984.

\end{thebibliography}

\end{document}